\RequirePackage{fixltx2e}
\documentclass[twocolumn,hyperpdf,amsmath,amssymb,aps,prd,10pt,superscriptaddress,nofootinbib,noeprint,preprintnumbers]{revtex4-1}

\usepackage{graphicx}
\usepackage{dcolumn}
\usepackage{bm}
\usepackage{amsmath}
\usepackage{amsfonts}
\usepackage{slashed}
\usepackage{hyperref}
\usepackage{multirow}
\usepackage{soul}
\usepackage{mathtools}
\usepackage{float}
\usepackage{afterpage}
\usepackage{color}
\usepackage{latexsym} 
\usepackage{rotating} 

\usepackage{marginnote}

\definecolor{jlab_red}{RGB}{192,39,45}
\definecolor{jlab_orange}{RGB}{249,102,0}
\definecolor{jlab_blue}{RGB}{47,122,121}
\definecolor{jlab_green}{RGB}{65,125,10}

\hypersetup{%
pdftitle = {Hadron Spectrum Collaboration},
pdfsubject = {QCD},
pdfkeywords = {QCD, Hadron, Physics, JLab, Lattice, Meson, Scattering},
pdfauthor = {Hadron Spectrum Collaboration},
colorlinks = {true},
filecolor = {black},
linkcolor = {jlab_green},
menucolor = {black},
citecolor = {jlab_green},
urlcolor = {jlab_green},
}{}

\newcommand{\cm}{\ensuremath{\mathsf{cm}}}

\newcommand{\kk}{\ensuremath{K\overline{K}}}
\newcommand{\kksm}{\ensuremath{K\!\bar{K}}}

\newcommand{\LG}{\mathrm{LG}}

\begin{document}

\preprint{JLAB-THY-15-2101}
\preprint{DAMTP-2015-34}

\title{Coupled $\pi\pi, K\overline{K}$ scattering in $P$-wave and the $\rho$ resonance from lattice QCD}


\author{David~J.~Wilson}
\email{djwilson@jlab.org}
\affiliation{Department of Physics, Old Dominion University, Norfolk, VA 23529, USA}

\author{Ra\'ul A. Brice\~{n}o}
\email{briceno@jlab.org}
\affiliation{Theory Center, Jefferson Lab, 12000 Jefferson Avenue, Newport News, VA 23606, USA}

\author{Jozef~J.~Dudek}
\email{dudek@jlab.org}
\affiliation{Theory Center, Jefferson Lab, 12000 Jefferson Avenue, Newport News, VA 23606, USA}
\affiliation{Department of Physics, Old Dominion University, Norfolk, VA 23529, USA}

\author{Robert~G.~Edwards}
\email{edwards@jlab.org}
\affiliation{Theory Center, Jefferson Lab, 12000 Jefferson Avenue, Newport News, VA 23606, USA}

\author{Christopher~E.~Thomas}
\email{c.e.thomas@damtp.cam.ac.uk}
\affiliation{Department of Applied Mathematics and Theoretical Physics, Centre for Mathematical Sciences, University of Cambridge, Wilberforce Road, Cambridge, CB3 0WA, UK}


\collaboration{for the Hadron Spectrum Collaboration}
\date{\today}

\begin{abstract}
We determine elastic and coupled-channel amplitudes for isospin-1 meson-meson scattering in $P$-wave, by calculating correlation functions using lattice QCD with light quark masses such that $m_\pi = 236$ MeV in a cubic volume of $\sim (4 \,\mathrm{fm})^3$. Variational analyses of large matrices of correlation functions computed using operator constructions resembling $\pi\pi$, $K\overline{K}$ and $q\bar{q}$, in several moving frames and several lattice irreducible representations, leads to discrete energy spectra from which scattering amplitudes are extracted. In the elastic $\pi\pi$ scattering region we obtain a detailed energy-dependence for the phase-shift, corresponding to a $\rho$ resonance, and we extend the analysis into the coupled-channel $K\overline{K}$ region for the first time, finding a small coupling between the channels.
\end{abstract}

\maketitle

The study of hadron spectroscopy from first principles QCD is entering a new stage of development where the relationship between the discrete spectrum of the theory in a finite-volume and the infinite-volume scattering amplitudes is being practically utilized to study resonances. The tool which allows us access to the spectrum is lattice QCD in which the quark and gluon fields are considered on a finite-grid of points with only systematically improvable approximations being made.

Predictably it is the simplest resonant scattering channel which has attracted the greatest initial interest~\cite{Aoki:2007rd,Feng:2010es,Lang:2011mn,Aoki:2011yj,Pelissier:2012pi}, that of $\pi\pi$ with isospin=1, in which a low-lying elastic vector resonance called the $\rho$ appears. These works have made use of the formalism relating the discrete spectrum at rest and in moving frames to {\it elastic} scattering amplitudes, which has been in place for many years~\cite{Luscher:1990ux,Rummukainen:1995vs,Kim:2005gf,Fu:2011xz,Leskovec:2012gb}. Recently the extension to coupled-channels has been presented~\cite{He:2005ey,Hansen:2012tf,Briceno:2012yi,Guo:2012hv}, and the first lattice QCD study of a coupled-channel system, that of $\pi K, \, \eta K$ in $S,P$ and $D$--waves, has appeared~\cite{Dudek:2014qha,Wilson:2014cna}, showing that the energy dependence and resonant content of the scattering matrix for such a system can be extracted from finite volume spectra.

To date virtually all determinations of hadron scattering amplitudes in lattice QCD calculations have worked with artificially heavy $u,d$ quark mass values, a choice which leads to heavier than physical pseudoscalar mesons -- this reduces the computational cost, allowing calculations in smaller volumes (where $m_\pi L$ remains large), and pushes up in energy the thresholds for multihadron scattering such as $\pi\pi\pi\pi$, for which a finite-volume formalism is not yet in place (but see Refs.~\cite{Hansen:2013dla, Meissner:2014dea, Hansen:2014eka,Hansen:2015zga} for progress).

The Hadron Spectrum Collaboration previously computed $\pi\pi$ scattering using 391 MeV pions~\cite{Dudek:2012gj,Dudek:2012xn}, extracting detailed spectra of QCD eigenstates from variational analysis of two-point correlation functions computed in several moving frames in three different volumes. By obtaining a significant number of energy levels in the elastic scattering region they were able to map out the energy dependence of the scattering amplitude and show that there is a narrow $\rho$ resonance barely above $\pi\pi$ threshold.

In this paper we deliver an extension of the work presented in~\cite{Dudek:2012xn}, utilizing a smaller $u,d$ quark mass, corresponding to a pion mass of $ 236\, \mathrm{MeV}$, in a large box of spatial extent $\sim 3.8\,\mathrm{fm}$. Going beyond what was done before, we also consider the effect of including $K\overline{K}$-like operators into the variational operator basis -- the enlarged basis allows us to determine energy levels above the $K\overline{K}$ threshold, and to extract first estimates within QCD of the coupled-channel $\pi\pi, K\overline{K}$ scattering matrix with $I=1$, $J^P=1^-$.

\section{Calculating the finite-volume spectrum} \label{sec_calc}

The results to be presented in this paper come from a calculation using a single ensemble of anisotropic Clover gauge-field configurations of volume ${(L/a_s)^3 \times (T/a_t) = 32^3 \times 256}$, with spatial lattice spacing $a_s \sim 0.12 \,\mathrm{fm}$, and temporal lattice spacing $a_t = a_s/\xi$ with $\xi \sim 3.5$. The $2+1$ flavors of dynamical quarks have strange quark mass tuned to approximate the physical strange quark~\cite{Edwards:2008ja,Lin:2008pr} and degenerate $u,d$ quarks with mass parameter $a_t m_\ell = -0.0860$ corresponding to a pion mass ${\sim 236\, \mathrm{MeV}}$ \cite{Babich:2011:SLQ:2063384.2063478,Joo:2012:LQG:2403996.2404003,Winter:2014:FLQ:2650283.2650646}. The large volume and time extent, $m_\pi L \sim 4.3$, and $m_\pi T \sim 10$, ensure that exponentially suppressed polarization and thermal effects will be negligible. Correlation functions are computed on 469 configurations, typically utilizing multiple time sources on each to increase statistics. 

Our approach is to determine the spectrum from a matrix of two-point correlations functions constructed using a basis of hadronic operators at source and sink. Our basis, which is described extensively in previous publications~\cite{Dudek:2009qf,Dudek:2010wm,Thomas:2011rh,Dudek:2012gj}, contains both ``single-meson-like'' operators of the form $\bar{\psi} \Gamma \overleftrightarrow{D} \ldots \overleftrightarrow{D} \psi$, and ``meson-meson-like'' operators of the form $\sum_{\vec{p}_1, \vec{p}_2} \mathcal{C}(\vec{p}_1, \vec{p}_2)\,  \Omega^\dag(\vec{p}_1)\,  \Omega^\dag(\vec{p}_2)$. $\Omega^\dag(\vec{p})$ is a variationally optimized combination of ``single-meson-like'' operators, capable of interpolating a stable pseudoscalar with momentum $\vec{p}$. The use of various meson momenta, $\vec{p}_1,\vec{p}_2$, at a fixed total momentum, $\vec{P} = \vec{p}_1+\vec{p}_2$, furnishes the ``meson-meson-like'' part of the operator basis, where the generalized Clebsch-Gordan coefficient, $\mathcal{C}(\vec{p}_1, \vec{p}_2)$ ensures the operator has the desired properties under rotations and parity.

In order to efficiently compute the correlation functions, whose Wick contractions include quark-antiquark annihilation on a timeslice, we make use of \emph{distillation}~\cite{Peardon:2009gh}. The distillation smearing operator is constructed as an outer product of the lowest 384 eigenvectors of the gauge-covariant laplacian on each timeslice. 
The light and strange quark propagators required to evaluate the correlation functions are the solutions of the Clover-Dirac equation using these eigenvectors as a source\footnote{The large number of propagators are very efficiently computed using an Adaptive Multi-Grid solver \cite{Osborn:2010mb,Babich:2010qb} for the light quarks on CPUs, and the strange quark propagators are efficiently computed using Graphical Processing Units \cite{Clark:2009wm,Babich:2010mu}.}. To our knowledge, this is the largest lattice volume on which the full distillation method has been applied.

We form the possible combinations of creation and annihilation operators at source and sink to construct a matrix of correlation functions. This can be analyzed variationally~\cite{Michael:1985ne,Luscher:1990ck,Blossier:2009kd} by solving a generalized eigenvalue problem, $C(t) v^\mathfrak{n} = \lambda_\mathfrak{n}(t) C(t_0) v^\mathfrak{n}$, where the eigenvalues $\lambda_\mathfrak{n}(t)$ give information about the spectrum, and the eigenvectors provide the optimal linear combination of basis operators to interpolate state $|\mathfrak{n}\rangle$. Details of our implementation may be found in Refs.~\cite{Dudek:2009qf,Dudek:2010wm}.

Through calculating the discrete spectrum in moving frames we may better constrain our description of scattering amplitudes -- each moving frame gives the boundary a different geometry, which lead to modified quantization conditions relating amplitudes to the spectrum.  In a finite, spatially periodic, cubic volume, there is a reduced symmetry with respect to that of an infinite space. For systems with no overall momentum, parity is a good quantum number and a double cover of the octahedral group, $O_h^D$, describes the symmetries of the system. The infinite volume partial waves are subduced into various irreducible representations, or ``irreps", of the octahedral group.
Systems may be considered having non-zero overall momentum, $\vec{P} = \frac{2\pi}{L}\big[n_x, n_y, n_z \big]$, which satisfy periodic boundary conditions if $n_x,n_y,n_z$ are integers. Such systems have a further reduced symmetry, relative to the rest-frame, and are described by the little groups, $\LG{(\vec{P})}$, which are the subgroups of the octahedral group whose transformations leave $\vec{P}$ unchanged. The irreps of $\LG{(\vec{P})}$ typically have an enlarged angular momentum content with respect to the system at rest, and parity is not a good quantum number. These concepts are discussed in detail in Ref.~\cite{Thomas:2011rh} with applications relevant to $\pi\pi$ scattering further developed in Refs.~\cite{Dudek:2012gj,Dudek:2012xn}.

\begin{figure}
\includegraphics[width=0.99\columnwidth]{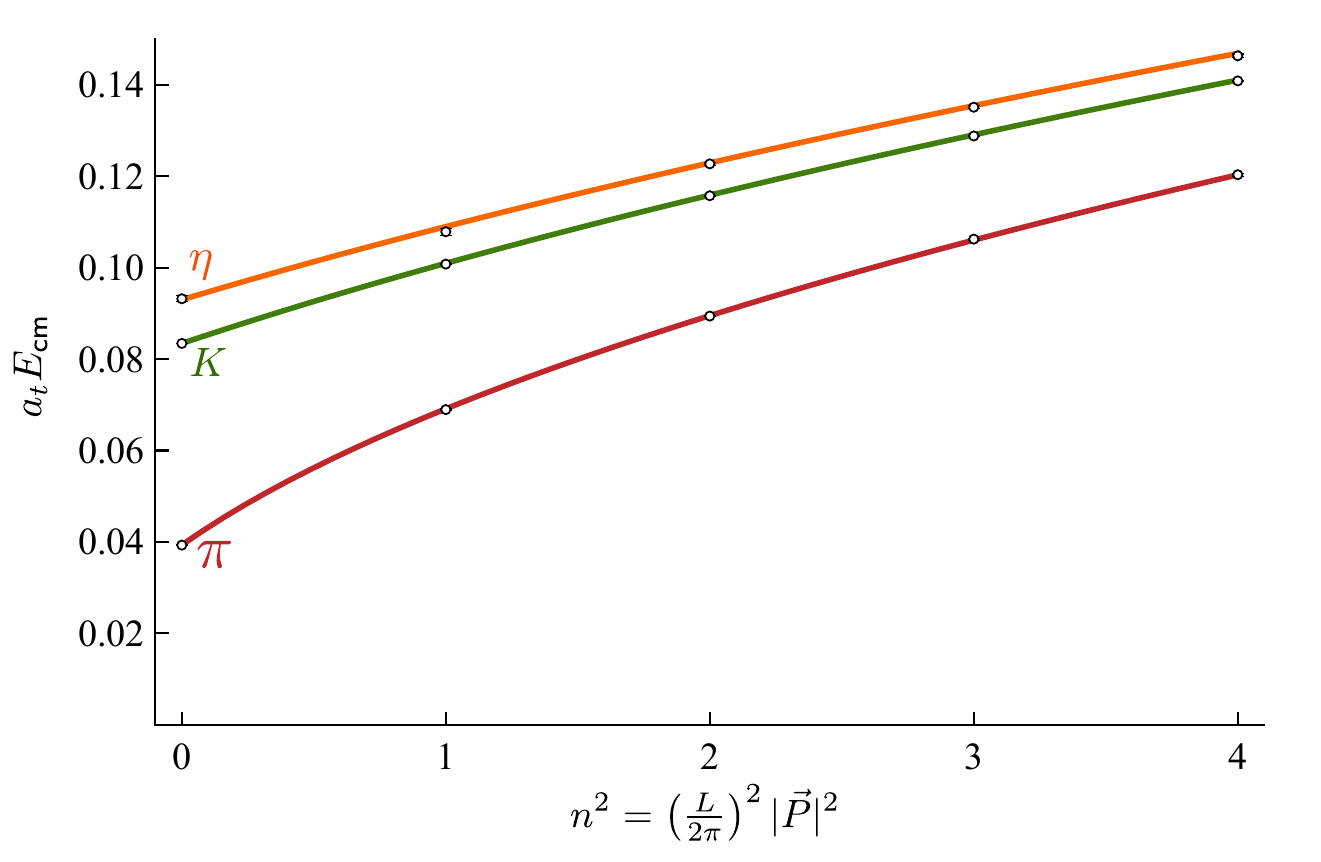}
\caption{Pseudoscalar $\pi$, $K$ and $\eta$ meson dispersion relations obtained on this lattice for $|\vec{P}|^2\left(\frac{L}{2\pi}\right)^2\le 4$ as described in the text. The statistical uncertainties on the energies are smaller than the symbols. The curves correspond to dispersion relation fits of the form $(a_t E)^2 = a_t^2 m^2 + \left(\frac{2\pi}{\xi L/a_s}\right)^2 n^2$ with parameters presented in Table~\ref{tab_masses_thresholds}.}
\label{fig_disp}
\end{figure}

The masses (and energies at nonzero momentum) of the stable pseudoscalars, $\pi, K$ and $\eta$, are obtained from the lightest eigenstates in the irreps $[000]\, A_1^+$ and ${|\vec{P}| > 0,\, A_2}$, and are presented in Figure~\ref{fig_disp}, along with dispersion relation fits of the form ${(a_t E)^2 = (a_t  m)^2 + \left(\frac{2\pi}{\xi L/a_s}\right)^2 n^2}$, where $\xi$ is allowed to vary for each pseudoscalar species -- Table~\ref{tab_masses_thresholds} shows the masses and extracted $\xi$ values. We observe a reasonable level of consistency in $\xi$ between species, and we choose to use the value from the pion, ${\xi = 3.453(6)}$, in the rest of this paper. The variational analysis that yields these energies also provides the eigenvectors used when forming the optimized operators $\Omega$ that are used in the ``meson-meson-like'' constructions.

\begin{table}
\begin{tabular}{c|c|c|c}
     & $a_t\, m$     & $\xi$ & $\chi^2/N_\mathrm{dof}$\\
\hline
$\pi$  & 0.03928(18) & 3.453(6)   & 1.4        \\
 $K$   & 0.08344(7)  & 3.462(4)   & 1.4        \\
$\eta$ & 0.09299(56) & 3.468(20)  & 0.73  
\end{tabular}
\begin{tabular}{c|c}
threshold       & $a_t E$\\ 
\hline
$\pi\pi$        & 0.07856(26) \\
$\pi\pi\pi\pi$  & 0.15712(52) \\
$\kk$           & 0.16688(14) \\
$\eta\pi\pi$    & 0.17155(62) \\
$\pi\kk$        & 0.20616(19) 
\end{tabular}
\caption{The masses of stable particles and anisotropy, $\xi$, obtained from dispersion fits using five momenta, as described in the text, and presented in Figure~\ref{fig_disp}. Threshold energies relevant to $I=1$, $J^P=1^-$ meson-meson scattering also presented. }
\label{tab_masses_thresholds}
\end{table}

\begin{figure*}
\includegraphics[width=0.99\textwidth]{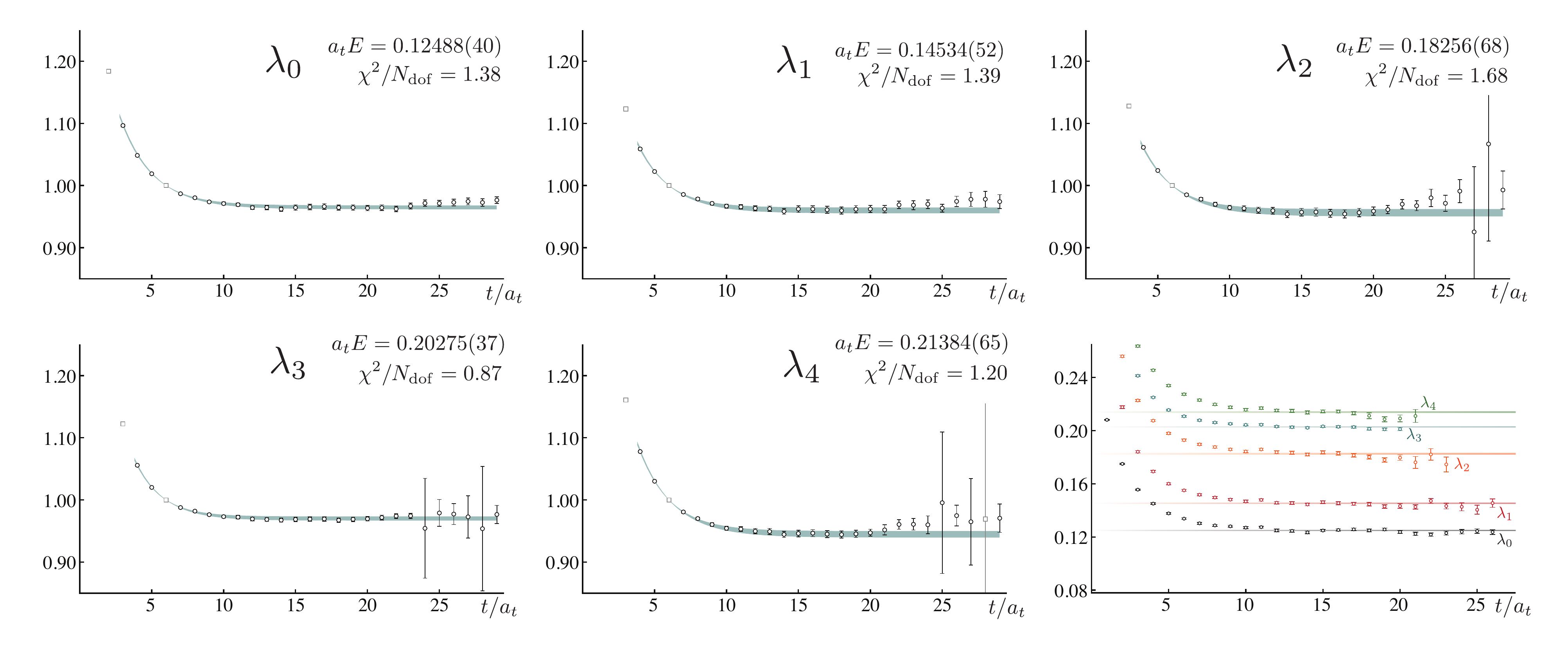}
\caption{Principal correlators, $\lambda_\mathfrak{n}(t)$, from the variational analysis of a $30\times 30$ matrix of $\vec{P}=[000]$, $\Lambda^P = T_1^-$ correlation functions, with $t_0 = 6 a_t$. The lowest five states are plotted as $e^{E_\mathfrak{n}(t-t_0)} \, \lambda_\mathfrak{n}(t)$, along with a two-exponential fit determining the energy, $E_\mathfrak{n}$. The bottom right panel shows the corresponding effective masses, $a_t m_\mathrm{eff} = \tfrac{1}{\delta t} \log \tfrac{\lambda_\mathfrak{n}(t)}{\lambda_\mathfrak{n}(t+ \delta t)}$ with $\delta t = 3 a_t$. \label{prin_corrs}}
\end{figure*}

Utilizing these methods, we computed $I=1$, $G$-parity positive spectra for all irreps\footnote{A table of the subductions of $\pi\pi$ (and equivalently $K\overline{K}$) partial-waves into these irreps appears as Table III in~\cite{Dudek:2012xn}.} containing $J^P=1^-$ with $|\vec{P}|^2 \le 4 \left(\tfrac{2\pi}{L}\right)^2$. The operator basis used to construct the correlation matrices for each lattice irrep, $\Lambda$, is presented in Table~\ref{tab_opsused}, where we see that the dimension of the correlator matrices to be considered is never smaller than 10 and may be as large as 37. As an example of the result of variational analysis, in Fig.~\ref{prin_corrs} we present the first five $\lambda_\mathfrak{n}(t)$ from the $\Lambda^P=T_1^-$ irrep which illustrates the quality of the spectra obtained. Higher energy levels are extracted, but are not used in the scattering analysis which follows.

\begin{table}
\begin{ruledtabular}
\begin{tabular}{ccccc}
$[000]\, T_1^-$     								& $[100]\, A_1$     							& $[110] \, A_1$ 					& $[111] \, A_1$				& $[200] \, A_1$ \\
\hline \hline
$\pi_{001} \,\pi_{00{\text -}1}$ 					& $\pi_{000} \,\pi_{100}$						& $\pi_{000} \,\pi_{110}$ 			& $\pi_{000} \,\pi_{111}$		& $\pi_{000} \, \pi_{200}$	\\	 
$\pi_{011} \,\pi_{0{\text -}1{\text -}1}$ 			& $\pi_{0{\text -}10} \,\pi_{110}$				& $\pi_{00{\text -}1} \,\pi_{111}$ 	& $\pi_{100} \,\pi_{011}$		& \\	 	
$\pi_{111} \,\pi_{{\text -}1{\text -}1{\text -}1}$ 	& $\pi_{0{\text -}1{\text -}1} \,\pi_{111}$		& $\pi_{{\text -}110} \,\pi_{200}$ 	& $\pi_{{\text -}111} \,\pi_{200}$		& \\	
													& $\pi_{{\text -}100} \,\pi_{200}$				&&&\\[1.5ex]
$K_{001} \,\overline{K}_{00{\text -}1}$ 			& $K_{000} \,\overline{K}_{100}$				& $K_{000} \,\overline{K}_{110}$ 	& $K_{000} \,\overline{K}_{111}$		& 	\\	
													& $K_{0{\text -}10} \,\overline{K}_{110}$		& $K_{00{\text -}1} \,\overline{K}_{111}$ 	& $K_{100} \,\overline{K}_{011}$		& 	\\[1.5ex]
$\bar{\psi}\mathbf{\Gamma} \psi  \times  26$		& $\bar{\psi}\mathbf{\Gamma} \psi  \times  10$	& $\bar{\psi}\mathbf{\Gamma} \psi  \times  13$ & $\bar{\psi}\mathbf{\Gamma} \psi  \times  21$ &  $\bar{\psi}\mathbf{\Gamma} \psi  \times  18$
\end{tabular}
\end{ruledtabular}

\vspace{.2cm}

\begin{ruledtabular}
\begin{tabular}{ccccc}
$[100]\, E_2$     									& $[110]\, B_1$     							& $[110] \, B_2$ 					& $[111] \, E_2$				& $[200] \, E_2$ \\
\hline \hline
$\pi_{0{\text-}10}\,\pi_{110}$						& $\pi_{010}\,\pi_{100}$						& $\pi_{00{\text-}1}\,\pi_{111}$ 	& $\pi_{100}\,\pi_{011}$		& $\pi_{1{\text-}10}\, \pi_{110}$ \\
$\pi_{0{\text-}1{\text-}1}\,\pi_{111}$				& $\pi_{01{\text-}1}\,\pi_{101}$				& $\pi_{01{\text-}1}\,\pi_{101}$	& $\pi_{{\text-}111}\,\pi_{200}$ & $\pi_{1{\text-}1{\text-}1}\, \pi_{111}$ \\
&													  $\pi_{{\text-}110}\,\pi_{200}$ \\[1.5ex]
$\bar{\psi}\mathbf{\Gamma} \psi  \times  29$		& $\bar{\psi}\mathbf{\Gamma} \psi  \times  29$	& $\bar{\psi}\mathbf{\Gamma} \psi  \times  29$ & $\bar{\psi}\mathbf{\Gamma} \psi  \times  35$ &  $\bar{\psi}\mathbf{\Gamma} \psi  \times  29$
\end{tabular}
\end{ruledtabular}

\vspace{.2cm}

\begin{ruledtabular}
\begin{tabular}{cccc}
&$[100]\, B_1$     									& $[100]\, B_2$  \\    							
\hline \hline
&$\pi_{0{\text-}10}\,\pi_{110}$						& $\pi_{0{\text-}1{\text-}1}\,\pi_{111}$ \\[1.5ex]	 
&$\bar{\psi}\mathbf{\Gamma} \psi  \times  9$		& $\bar{\psi}\mathbf{\Gamma} \psi  \times  9$						
\end{tabular}
\end{ruledtabular}

\caption{The operator bases used in each lattice irrep in this calculation. For each irrep we list the ``$\pi\pi$-like'' and ``$\kk$-like'' operators (Appendix \ref{app:ops} contains details of the $K\overline{K}$ operator construction) that were used as well as the number of ``single-meson-like'' operators. We use a notation which indicates the momentum (in units of $2\pi/L$) of the pseudoscalar meson operators, recalling that the directions of momentum are summed over with generalized Clebsch-Gordan weights to ensure the operator lies in the stated irrep~\cite{Dudek:2012gj,Dudek:2012xn}.}
\label{tab_opsused}
\end{table}

\begin{figure*}
\includegraphics[width=0.99\textwidth]{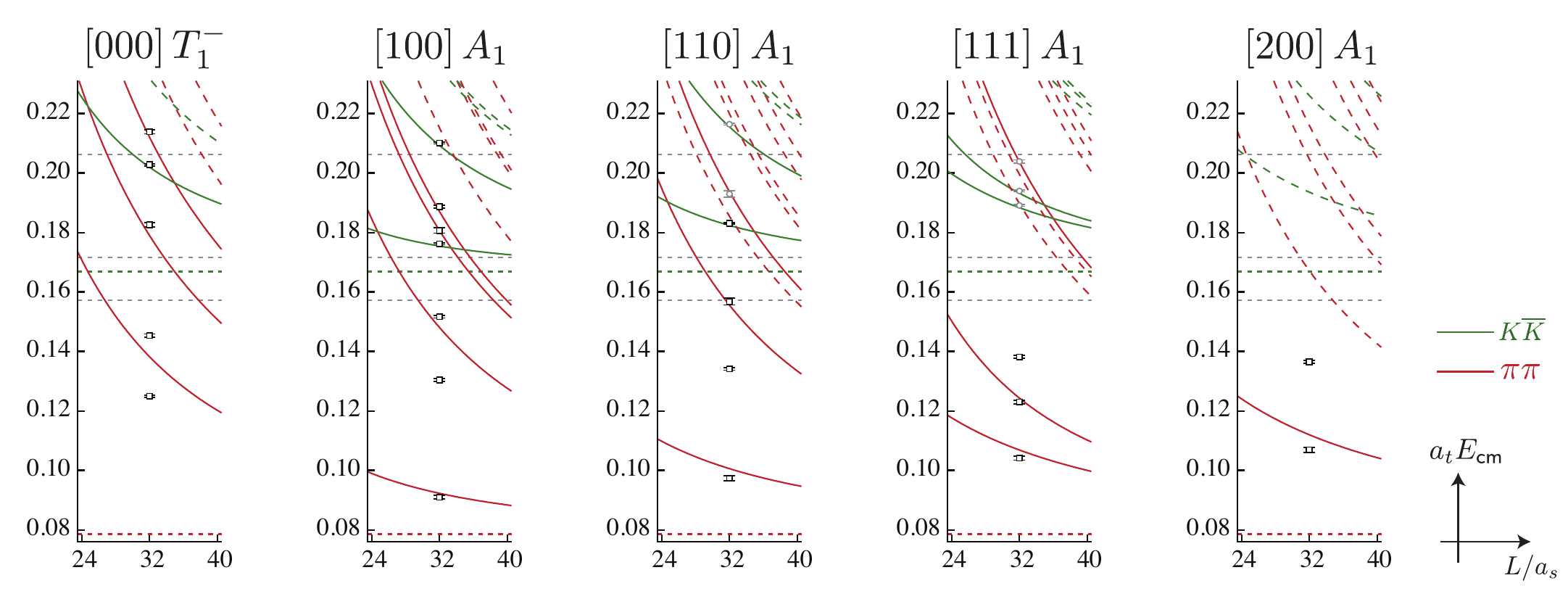}
\caption{Finite volume spectra obtained in the $T_1^-$ and moving frame $A_1$ irreps. Short dashed lines denote meson-meson and multimeson thresholds, with $\pi\pi$ in red and $\kk$ in green. Solid curves are allowed non-interacting meson-meson energies in finite volume, corresponding to operator constructions that were included in our basis, while the long-dashed curves are those that were not included in our basis. The points show the energy levels with their statistical errors as extracted from the lattice QCD correlation functions, with those in black being the ones used in the amplitude analysis to follow and those in grey not used, as described in the text.}\label{spec_T1_A1}
\end{figure*}

\begin{figure*}
\includegraphics[width=0.99\textwidth]{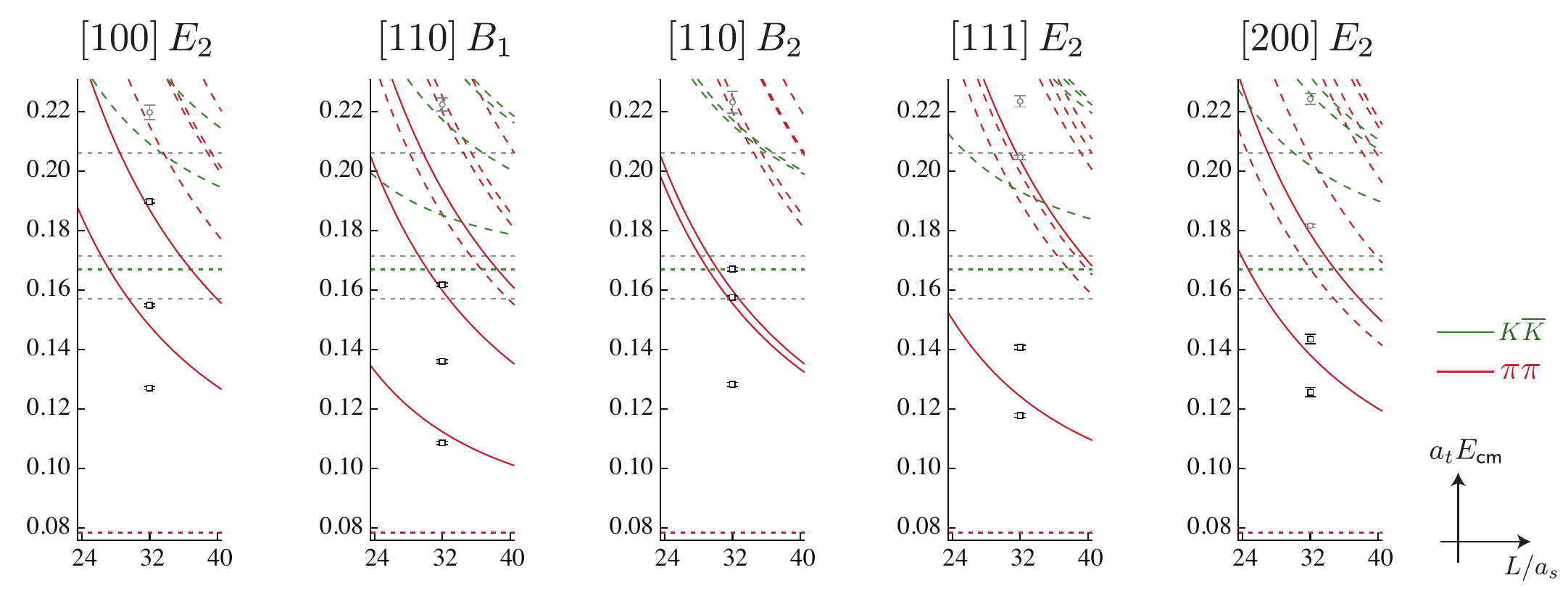}
\caption{As Fig.~\ref{spec_T1_A1} for the $E$- and $B$-type little-group irreps which have $J^P = 1^-$ as the lowest subduced partial wave.}\label{spec_E_B}
\end{figure*}

\begin{figure}
\includegraphics[width=0.90\columnwidth]{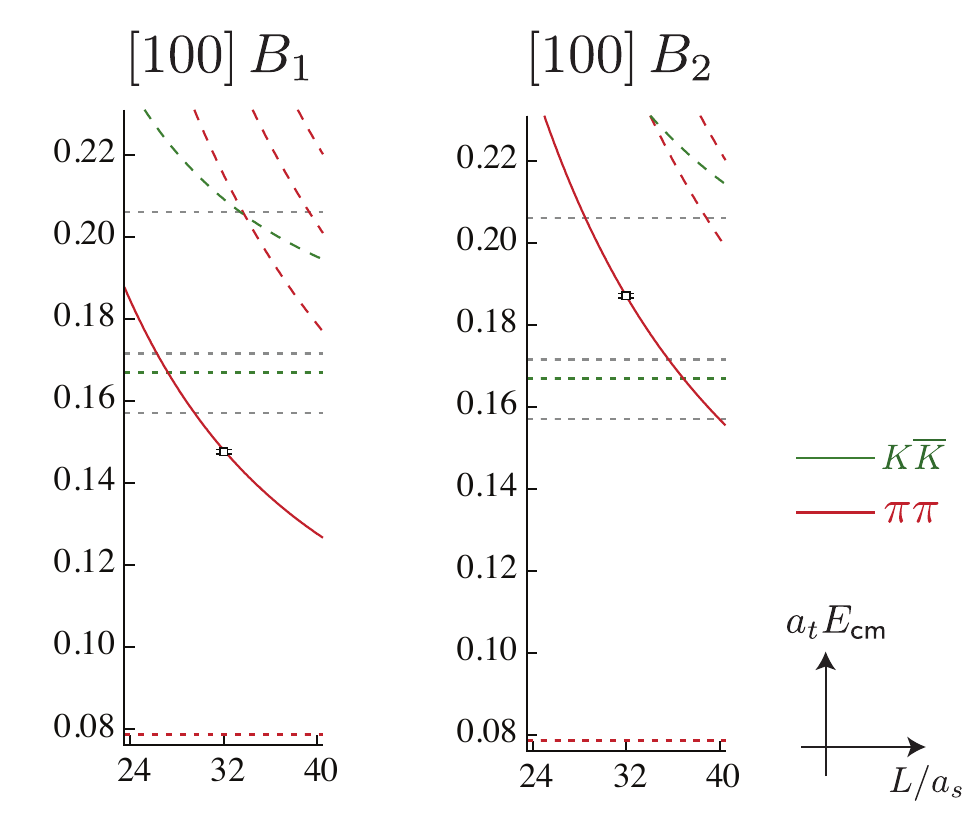}
\caption{As Fig.~\ref{spec_T1_A1} for the $B$-type little group irreps with $\vec{P} = [100]$ which have $J^P = 3^-$ as the lowest subduced partial wave.}\label{spec_J3}
\end{figure}

The spectra extracted are shown in Figures~\ref{spec_T1_A1}, \ref{spec_E_B} and \ref{spec_J3}, where we display the energy levels transformed to their $\mathsf{cm}$-frame values (points with statistical error bars), the scattering thresholds (short-dashed horizontal lines) and the spectrum of non-interacting $\pi\pi$ and $K\overline{K}$ levels (curves). As can be seen in Table~\ref{tab_opsused}, we have not included any operators featuring pseudoscalar operators with momentum, $|\vec{p}|^2 > 4$. The long dashed curves indicate those non-interacting levels for which we have not included the corresponding operator construction. The high-lying extracted levels displayed by gray circles lie in an energy region in which we have not included sufficient operators to reliably extract the entire spectrum, and these levels will not be used in the analysis which follows. While the $A_1$ little-group irreps contain no subductions from the $J^P=1^+$ partial-wave, the $E$ and $B$ irreps do, and this is likely to be the origin of the consistent ``additional'' level, present near $a_t E_\mathsf{cm} \sim 0.22$ in each pane of Figure~\ref{spec_E_B}, being due to a positive parity $b_1$ resonance. These states are observed to have large overlap onto ``single-meson'' operators subduced from $J^P=1^+$. 

The light gray dashed horizontal lines in Figures~\ref{spec_T1_A1}, \ref{spec_E_B} and \ref{spec_J3} show the multihadron thresholds, $4\pi$, $\eta \pi\pi$ and $\pi K\overline{K}$. Note that we have not included operators resembling these in our variational basis, nor have we plotted the corresponding non-interacting levels in the Figures -- these will lie at a higher energy than the threshold. Experimentally these channels have very small amplitudes in $e^+ e^-$ annihilation until several hundred MeV above threshold~\cite{Aubert:2005eg,Aubert:2007ef}, so we do not expect them to play a significant role -- we will discuss this in greater detail later in this paper.

\begin{figure*}
\includegraphics[width=0.75\textwidth]{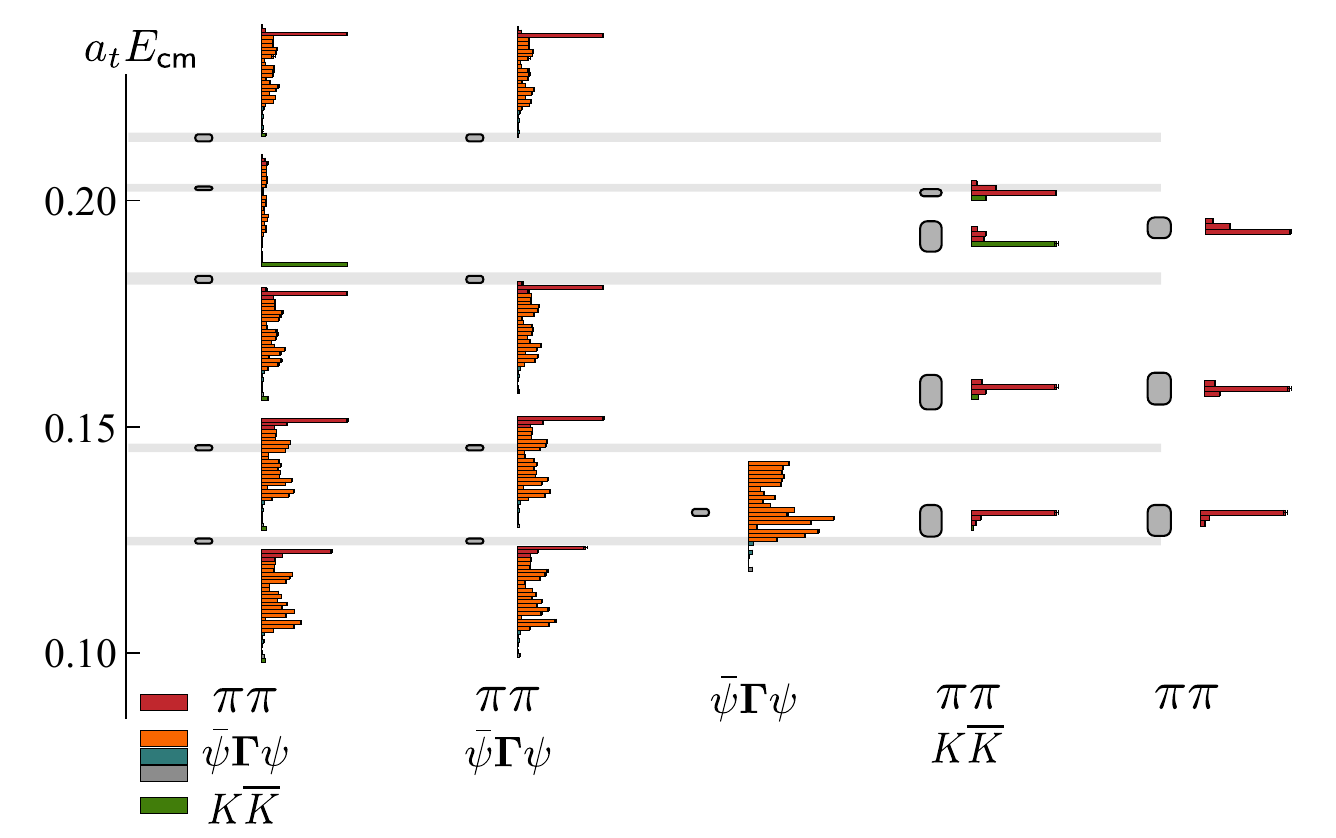}
\caption{The spectra obtained from various choices of operator basis in the $T_1^-$ lattice irrep. The leftmost column contains all of the operators we considered, including ``single-meson-like'' operators subduced from $J=1$(orange), $J=3$(blue) and $J=4$(grey). The remaining columns use fewer operators as indicated. The histograms show the suitably normalized magnitudes of the contributions of each operator to each energy level, $\langle \mathfrak{n} | \mathcal{O}^\dag | 0 \rangle$, obtained from the variational solutions. See \cite{Dudek:2009qf, Dudek:2010wm} for more details. }
\label{op_basis}
\end{figure*}

Before attempting to determine meson-meson scattering amplitudes from the spectra presented in Figures~\ref{spec_T1_A1}, \ref{spec_E_B} and \ref{spec_J3}, we will present a brief illustration of the importance of using a sufficiently diverse basis of operators in variational analysis. Figure~\ref{op_basis} shows the spectrum extracted in the $[000]\, T_1^-$ irrep using five different choices of operator basis. The histograms show the relative strength of overlap $\langle \mathfrak{n} | \mathcal{O}^\dag |0 \rangle$ for the various operators in the basis\footnote{see \cite{Dudek:2009qf, Dudek:2010wm} for further details of the normalization of such overlaps}. The leftmost column is our largest basis, the one presented in Table~\ref{tab_opsused} which contains three $\pi\pi$-like operators, one $K\overline{K}$-like operator, and 26 ``single-meson-like'' operators, 19 of which are subduced from $J^P=1^-$ constructions, 6 from $J^P=3^-$ subductions and one from a $J^P=4^-$ subduction. The second column lacks the $K\overline{K}$-like operator and is seen to give a reasonably consistent spectrum with the exception of the level which had large overlap onto the $K\overline{K}$-like operator. The third column uses only the ``single-meson-like" operators, lacking any $\pi\pi$-like or $K\overline{K}$-like constructions -- the only low lying state extracted appears to be some sort of crude average of the two lowest lying states. The fourth and fifth columns, which exclude ``single-hadron-like'' operators also provide poor determinations of the spectrum. It appears, as one might expect for a system in which we expect a narrow resonance, usually thought of as a tightly bound $q\bar{q}$ state, strongly coupled to $\pi\pi$, an accurate spectrum cannot be determined without including both ``single-meson-like'' operators and $\pi\pi$-like operators. A simple argument explaining these observations (illustrated using an in-flight irrep) was previously given in~\cite{Dudek:2012xn}. In Appendix \ref{app:single} we discuss the result of performing a phase-shift extraction using the spectrum extracted using only ``single-meson-like'' operators.

We briefly comment that in the $[000]\, T_1^-$ irrep, there are extracted levels near $a_t E_\mathsf{cm} \sim 0.28$ which have significant overlap onto the $\bar{\psi}\mathbf{\Gamma} \psi$ operators, these likely indicate the mass scale of the higher excited vector resonances. We also find levels that have significant overlaps with operators subduced from continuum $J^P=3^-$ and $J^P=4^-$ above $a_t E_\cm=0.33$, suggesting $\rho_3$, $\rho_4$ resonances.

\section{Determining scattering amplitudes} \label{sec_amps}

The relationship between a two-body coupled-channel scattering $t$-matrix and the discrete spectrum for an irrep $\vec{P},\Lambda$ in a finite, periodic, $L\times L \times L$ volume is provided by the equation,
\begin{align}
\det\Big[&
\delta_{ij}\delta_{\ell\ell^\prime}\delta_{nn^\prime}  \nonumber \\
&\;+ i\rho^{}_i(E_\mathsf{cm})\, t^{(\ell)}_{i j}(E_\mathsf{cm})
\left(\delta_{\ell\ell^\prime} \delta_{nn^\prime}+
i  \mathcal{M}^{\vec{P},\Lambda}_{\ell n;\ell^\prime n^\prime}(q^2_i) \right)
\Big] = 0.
\label{eq_luescher_t}
\end{align}
where the determinant is over the channel indices $i$ and the partial-waves, $\ell$, subduced into irrep $\Lambda$. ${\rho^{}_i(E_\mathsf{cm}) = 2 k_i / E_\mathsf{cm}}$ is the phase space for channel $i$, and the finite-volume dependent matrix $\mathcal{M}^{\vec{P},\Lambda}_{\ell n;\ell^\prime n^\prime}$, with ${q_i=k_i L/2\pi}$ where $k_i$ is the $\cm$ momentum in channel $i$, is described in Eq.~7 of Ref.~\cite{Wilson:2014cna} and Eq.~89 of Ref.~\cite{Rummukainen:1995vs}. This expression was derived in Refs.~\cite{He:2005ey,Hansen:2012tf,Briceno:2012yi,Guo:2012hv}, and in the case of a single open channel, reduces to the conditions presented earlier in~\cite{Luscher:1990ux} and~\cite{Rummukainen:1995vs,Kim:2005gf}. In the elastic case, $t^{(\ell)} = \tfrac{1}{\rho} e^{i \delta_\ell} \sin \delta_\ell$, and scattering can be described by a single real function, the scattering phase-shift, $\delta_\ell(E_\mathsf{cm})$.

For a given $t$-matrix, the solutions of Eq.~\ref{eq_luescher_t} provide the finite volume spectrum, $\{E_\mathfrak{n}\}$, in each lattice irrep $\Lambda$ with some overall momentum $\vec{P}$. In the elastic case, if higher partial-waves have negligibly small amplitudes, as one expects at low energies, the equation can be solved for $\delta_1(E_\mathfrak{n})$ for each calculated $E_\mathfrak{n}$. In a two-channel scattering problem there are three unknown functions of energy to determine for each partial-wave\footnote{three independent parameters are required to describe a unitary, time-reversal invariant, two-channel $t$-matrix} so for a given level $E_\mathfrak{n}$ this equation is underconstrained. If higher partial waves are not negligible, there will be still further unknowns. Fortunately, we are able to extract multiple energy levels in many irreps and these can be simultaneously used to constrain the scattering amplitude as a function of energy. By parameterizing the energy-dependence of the $t$-matrix, we can minimise a $\chi^2$ function describing the difference between the calculated spectrum and the spectrum given by solutions of Eq.~\ref{eq_luescher_t} for the $t$-matrix parameterization\footnote{The explicit form of the $\chi^2$ is provided in Eq.~9 of ref.~\cite{Dudek:2012xn}}. This method was first applied to a coupled-channel situation using lattice QCD energy levels in Ref.~\cite{Dudek:2014qha} and further details of this method and our implementation may be found in Ref.~\cite{Wilson:2014cna}.

\subsection{Elastic $\pi\pi$ scattering} \label{sec_elastic}

\begin{figure*}
\includegraphics[width=0.75\textwidth]{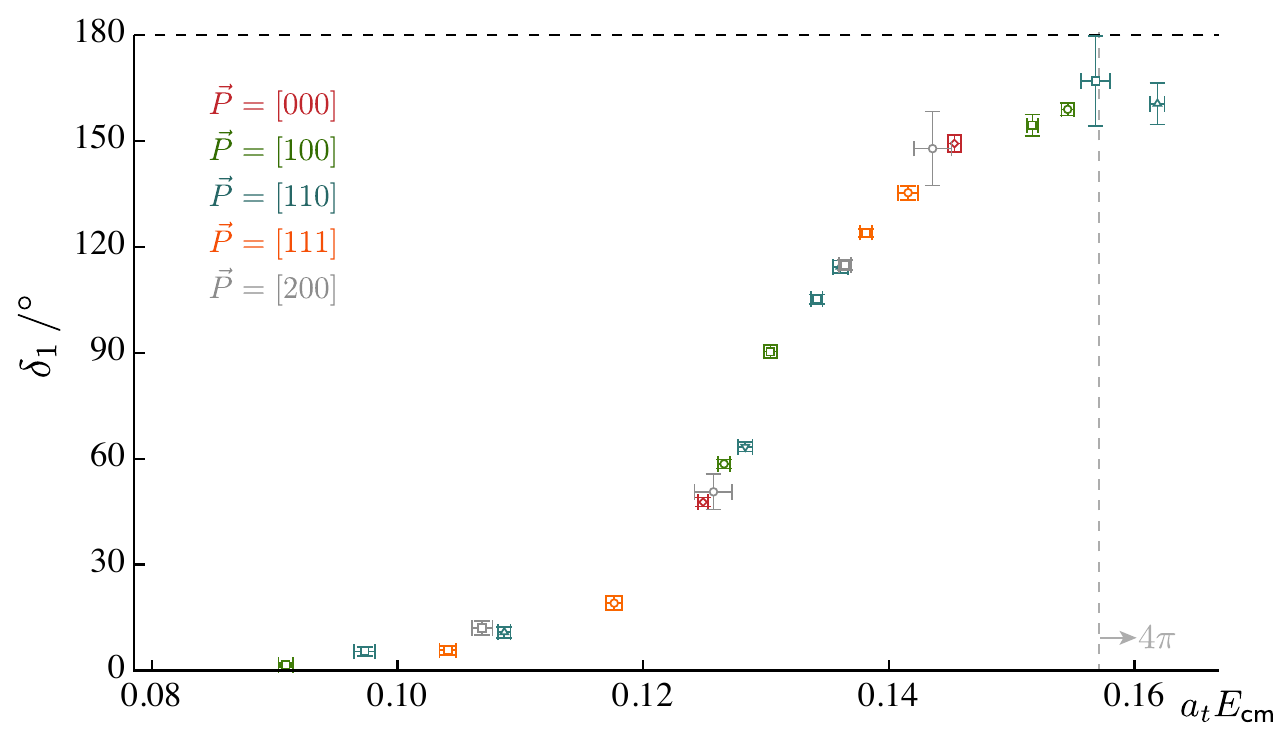}
\caption{$P$-wave $\pi\pi$ isospin-1 elastic scattering phase-shift extracted from energy levels plotted in Figures~\ref{spec_T1_A1} and \ref{spec_E_B}, assuming $F$-wave and higher partial-wave amplitudes are negligible in this energy region. Two points whose phase-shift values have rather large error bars are not shown. Grey dashed vertical line shows the $\pi\pi\pi\pi$ threshold.  
}\label{elastic_points}
\end{figure*}

We first study the elastic region, by considering only those levels extracted below the $4\pi$ threshold, which lies slightly below the $K\overline{K}$ threshold. We will initially proceed assuming that only $\pi\pi$ scattering in $P$-wave is relevant in this energy region, and later show that indeed the $\pi\pi$ $F$-wave amplitude and higher play no significant role. When partial waves above $\ell=1$ are negligible, then using Eq.~\ref{eq_luescher_t} one can obtain an estimate of $\delta_1(E_\mathsf{cm})$ at each calculated value of $E_\mathsf{cm}$, as plotted in Figures~\ref{spec_T1_A1} and \ref{spec_E_B}. These phase-shift values are plotted in Figure~\ref{elastic_points}, where we see a clear resonant behavior above $\pi\pi$ threshold. 

In order to describe the resonant content of the scattering amplitude we may explore energy-dependent parameterizations. We will consider various choices of energy-dependent parameterization in the $\chi^2$ minimization described above and will later discuss their pole content, finding that all choices capable of describing the finite volume spectrum have a pole at the same location in the complex-energy plane, corresponding to a single resonance.

\begin{figure}
\includegraphics[width=0.95\columnwidth]{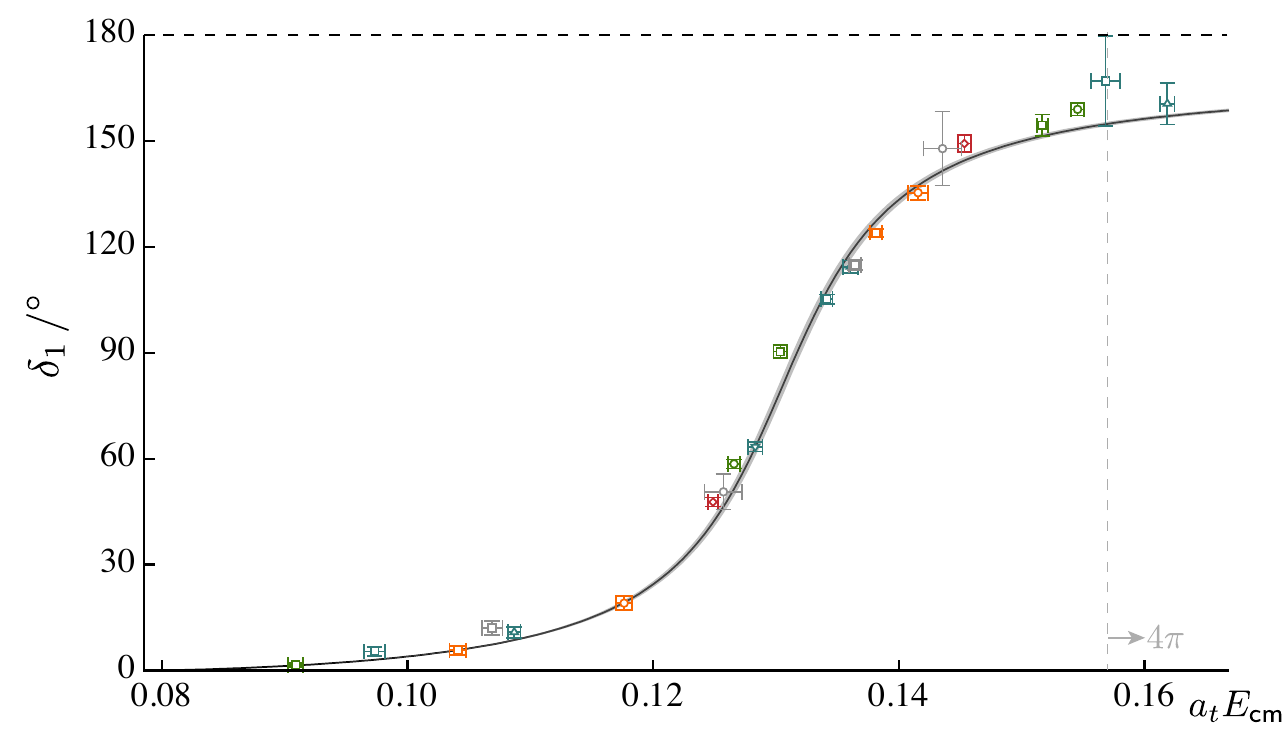}
\caption{$P$-wave $\pi\pi$ isospin-1 elastic scattering phase-shift. Points as in Figure~\ref{elastic_points}. Curve shows the Breit-Wigner description whose parameters are given in Eq.~\ref{bw_par_values}.\label{elastic_points_BW}}
\end{figure}

In elastic scattering, the Breit-Wigner parameterization is commonly used to describe isolated resonances -- in our case, with $s=E_\mathsf{cm}^2$, this takes the form
\begin{equation}
t(s) = \frac{1}{\rho(s)}\,  \frac{\sqrt{s}\,  \Gamma(s)}{m_R^2 -s - i \sqrt{s} \, \Gamma(s)},  \label{BW}
\end{equation}
with the energy dependent width, $\Gamma(s) = \frac{g_R^2}{6\pi} \frac{k^3}{s}$, including a coupling constant, $g_R$, and the threshold behavior required in $P$-wave scattering. Attempting to describe 22 levels below $4\pi$ threshold, we find the following parameters,

\vspace{0.2cm}
\begin{tabular}{rll}
$m_R=$ & $ 0.13171\, (36)\, (6) \cdot a_t^{-1}$ & \multirow{2}{*}{ $\begin{bmatrix*}[r] 1 & 0.04 \\ & 1\end{bmatrix*}$ } \\
$g_R=$ & $ 5.691 \, (70) \, (25)$   & \\[1.3ex]
&\multicolumn{2}{l}{ $\chi^2/ N_\mathrm{dof} = \frac{24.92}{22 - 2} = 1.25   \;, $}  \\
\end{tabular}
\vspace{-0.8cm}
\begin{equation} \label{bw_par_values}\end{equation}
\noindent
where the first set of errors describes the statistical uncertainty and the second comes from varying the pion mass and anisotropy, $\xi$, within their uncertainties. The matrix illustrates the statistical correlation between parameters, which in this case is seen to be very small. The corresponding $\delta_1(E_\mathsf{cm})$ is plotted in Figure~\ref{elastic_points_BW}.

\begin{figure}
\includegraphics[width=0.95\columnwidth]{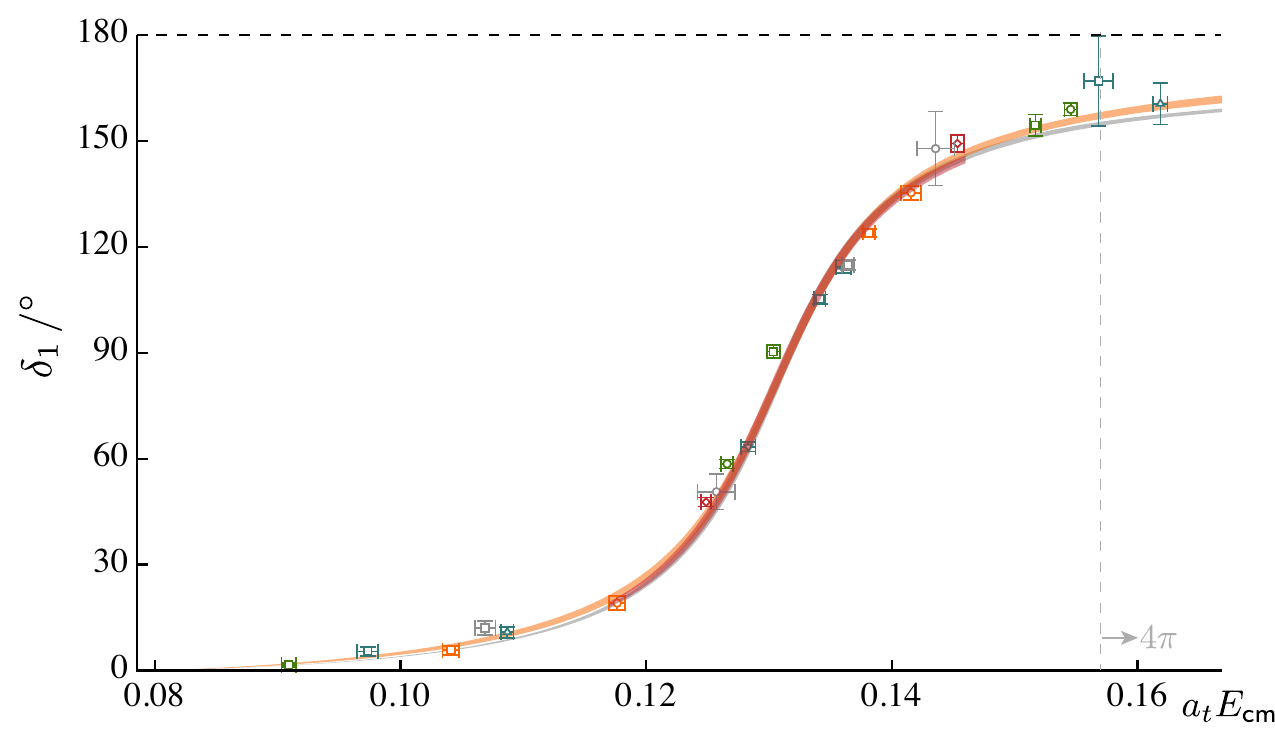}
\caption{$P$-wave $\pi\pi$ isospin-1 elastic scattering phase-shift. Points as in Figure~\ref{elastic_points}. Curves show the Breit-Wigner fit to the whole elastic region (grey), a Breit-Wigner with Hippel-Quigg~\cite{VonHippel:1972fg} barrier corrections (orange), and a Breit-Wigner description of a narrower energy region around the resonance peak (red).}\label{elastic_points_BW_variations}
\end{figure}

Modifications to the Breit-Wigner form which tame the $k^3$ barrier behavior at higher energies can be considered~\cite{Dudek:2012xn,VonHippel:1972fg} -- fits to 22 levels with these forms yield barely improved $\chi^2$ values and values of $m_R$ and $g_R$ that are statistically compatible with those given above. Restricting the energy region being described by the Breit-Wigner of Eq.~\ref{BW} to $0.117 < a_t E_\mathsf{cm} < 0.146$, i.e. excluding the tails of the resonance, leaves 14 energy levels -- fitting these also leads to $m_R, g_R$ values compatible with those given above. The corresponding phase-shifts for these modified fits are plotted in Figure~\ref{elastic_points_BW_variations}.

A more flexible parameterization scheme is provided by the $K$-matrix, which automatically satisfies unitarity in the single-channel and coupled-channel cases. Our implementation is presented in~\cite{Wilson:2014cna} and reads, for $\ell$-wave scattering,
\begin{align}
t_{ij}^{-1}(s) = \frac{1}{(2k_i)^\ell}K^{-1}_{ij}(s) \frac{1}{(2k_j)^\ell} + I_{ij}(s)\,,
\label{eq_t_kmat}
\end{align}
where $K_{ij}(s)$ is a real function, and $I_{ij}(s)$ is the Chew-Mandelstam function whose imaginary part above thresholds, $\mathrm{Im}\, I_{ij}(s) = - \delta_{ij} \,\rho_i(s)$, ensures unitarity, and whose real part is defined by a dispersive integral that ensures that $t(s)$ has no pseudothreshold branch point (at $s=0$).

\begin{table*}
\begin{ruledtabular}
\begin{tabular}{c ccccc l}
$N$ & $a_t m$   &$g$         & $\gamma_0/a_t^2$ & $\gamma_1/a_t^2$ & $\gamma_2/a_t^2$ & $\chi^2/N_\mathrm{dof}$ \\
\hline\hline\\[-2.0ex]
--& 0.13172(36) & 0.4475(52) & --              & --         & --           & $27.0/(22 - 2) =  1.3$\\
0 & 0.13164(36) & 0.4611(66) & $5.4(17)$       & --         & --           & $16.8/(22 - 3) =  0.88$\\
1 & 0.13161(37) & 0.4677(82) & $-3.3(67)$      & $2.6(22)$  & --           & $15.6/(22 - 4) =  0.86$\\
2 & 0.13165(37) & 0.4679(89) & $-21.5(74)$     & $16.6(24)$ & $-2.4(4)$    & $14.8/(22 - 5) =  0.87$ 
\end{tabular}
\end{ruledtabular}
\caption{$K$-matrix descriptions of the elastic spectrum using Eq.~\ref{eq_kmat_elastic}.}
\label{tab_Kmatrix_elastic}
\end{table*}

\begin{figure}
\includegraphics[width=0.95\columnwidth]{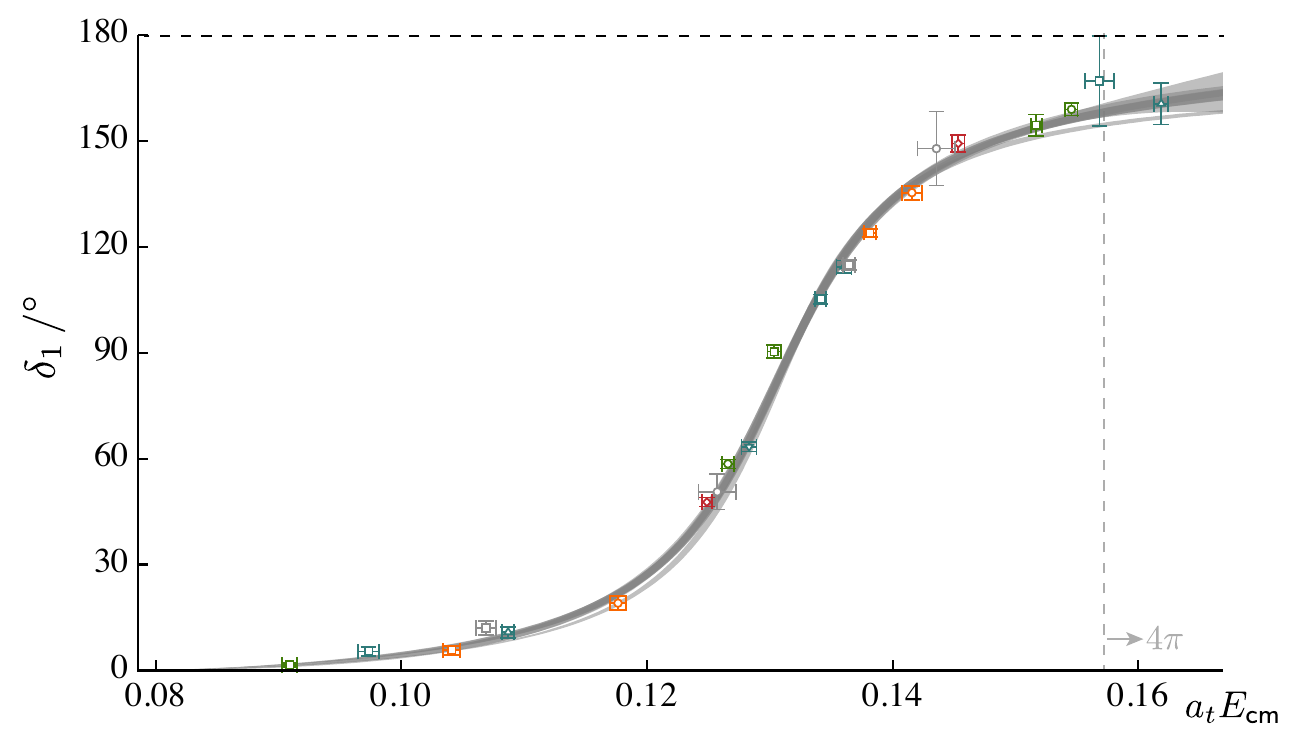}
\caption{$P$-wave $\pi\pi$ isospin-1 elastic scattering phase-shift. Points as in Figure~\ref{elastic_points}. Curves illustrate the $K$-matrix descriptions given in Table~\ref{tab_Kmatrix_elastic}, plus a $K$-matrix description using the simple phase-space, $\rho$, in place of the Chew-Mandelstam function. }\label{elastic_points_K_variations}
\end{figure}

In single-channel $\pi\pi$ scattering with $\ell=1$, the $K$-matrix is just a single function, and a convenient form is 
\begin{equation}
K(s) = \frac{g^2}{m^2 - s} + \sum_{n=0}^N \gamma_n \, \left(\frac{s}{s_0}\right)^n,
\label{eq_kmat_elastic}	
\end{equation}
with $s_0 = 4m_\pi^2$. Along with a suitable subtraction in the dispersive integral for $I(s)$ so that ${\mathrm{Re}\, I(s=m^2) = 0}$, this gives a $t(s)$ behavior that is similar to a Breit-Wigner, but with the polynomial allowing more freedom in the energy dependence. The 22 energy levels below $4\pi$ threshold have been described by this form for three choices, $N=0,1,2$, and without any polynomial term at all -- the results are presented in Table~\ref{tab_Kmatrix_elastic}. There is negligible improvement in the $\chi^2/N_{\mathrm{dof}}$ adding terms linear or quadratic in $s$, and the corresponding parameters are found to possess an increasingly large degree of correlation. The phase-shifts corresponding to the fits in Table~\ref{tab_Kmatrix_elastic} are plotted in Figure~\ref{elastic_points_K_variations}.

The assumption that $\pi\pi$ $F$-wave scattering plays a negligible role in determining the spectrum in the elastic region, as was assumed in the previous analysis, can be tested using the energy levels we have determined. The irreps $[100]$ $B_1$ and $B_2$ have $J^P = 3^-$ as their leading partial-wave, and we have determined one low-lying energy level in each, as shown in Fig.~\ref{spec_J3}. Using Eq.~\ref{eq_luescher_t} to obtain the phase-shift, under the reasonable assumption that higher partial-waves are negligible, we find $\delta^{\pi\pi}_3=(0.45 \pm 0.60 \pm 0.28)^\circ$ from the point at ${a_t E_\mathsf{cm} = 0.14763(51)}$ in $[100]B_1$. The level at $a_t E_\cm= 0.18712(53)$ in $[100]B_2$, which is slightly outside the elastic region, yields a phase-shift of $\delta^{\pi\pi}_3=(-0.2 \pm 1.6 \pm1.8)^\circ$. It is clear that the $F$-wave is negligibly small at low energies.

We may repeat the analysis described above to determine the $P$-wave scattering amplitude, also allowing a non-zero $F$-wave amplitude to influence the spectrum. A description of the 22 levels described previously plus the $[100]B_1$ level, using a Breit-Wigner to describe the $P$-wave and a scattering length parameterization for the $F$-wave, $k^7 \cot \delta_3=1/a_3$, yields $a_3=19(14)\times 10^5\,a_t^7$, or $m_\pi^7 \cdot a_3 = 27(20)\times 10^{-5}$, and $P$-wave Breit-Wigner parameters statistically compatible with those given above. $K$-matrix variations produce similar results, with the $F$-wave amplitude always being compatible with zero.

\subsection{Coupled-channel $\pi\pi$, $K\overline{K}$ scattering} \label{sec_coupled}

We now consider the coupled-channel region above $K\overline{K}$ threshold, where $\pi\pi \to K \overline{K}$ is expected to be the first significant source of inelasticity. Although we will use levels which lie above the $4\pi$ and $\eta \pi \pi$ thresholds, we will not consider those to be open channels. We expect the scattering amplitudes featuring these channels to be very small in the near-theshold energy region -- experimental support for this assertion comes from the measured cross-sections for $e^+ e^- \to 4\pi$~\cite{Aubert:2005eg} and $e^+ e^- \to \eta \pi \pi$~\cite{Aubert:2007ef}, neither of which has any significant value until at least 300 MeV above threshold, likely due to the dominance of meson-meson isobars in the amplitudes. Our expectation is that if we were to include operators resembling $4\pi$ and/or $\eta \pi \pi$ into our basis, we would extract additional energy levels very close to non-interacting levels corresponding to weak scattering amplitudes, decoupled from the $\pi \pi$, $K\overline{K}$ channels that we consider. These non-interacting levels will lie somewhat above the corresponding thresholds. The formalism to understand three-body and higher multiplicity scattering amplitudes is not yet complete, although recent progress is promising~\cite{Hansen:2013dla,Hansen:2014eka,Hansen:2015zga}.

We consider coupled-channel $K$-matrices like those described in Ref.~\cite{Wilson:2014cna}, using Eq.~\ref{eq_t_kmat} to define the $t$-matrix and $K_{ij}$ being a $2\times 2$ matrix. A particularly useful form for $K$ is
\vspace{-0.4cm}
\begin{align}
K_{ij}(s)=\frac{g_i \, g_j}{m^2-s} + \sum_{n=0}^N \gamma^{(n)}_{ij}  \left(\frac{s}{s_0}\right)^n,
\label{eq_km_cc}
\end{align}
\vspace{-0.4cm}

\noindent where the explicit pole in the first term is an efficient way of obtaining a coupled-channel pole in the $t$-matrix. While this parameterization permits a pole to occur in the complex energy plane, it is the description of the finite volume energy levels which determines whether or not this pole occurs close to the real axis and is thus relevant. We use the Chew-Mandelstam form for the phase space, subtracted at the pole position so that ${\mathrm{Re}\, I_i(s=m^2) = 0}$.

\begin{figure}[h]
\includegraphics[width=0.95\columnwidth]{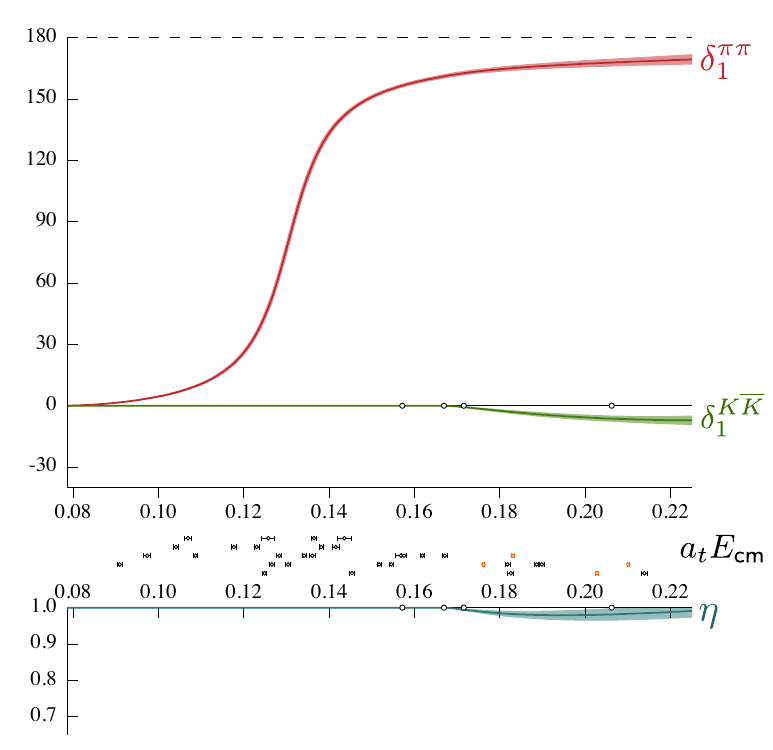}
\caption{Coupled $P$-wave $\pi\pi$ and $K\overline{K}$ isospin-1 phase shifts $\delta$ and inelasticity $\eta$ from a single $K$-matrix fit. Statistical uncertainty shown by the shaded band. The central points show the energy levels constraining the amplitude extraction with those dominated by $\pi\pi$-like and $\overline{q}q$-like operators shown in black and those with significant $K\overline{K}$ contributions shown in orange. On axis circles show the opening of the $4\pi$, $\kk$, $\eta \pi\pi$ and $\pi \kk$ thresholds.}\label{coupled_default_K}
\end{figure}

\begin{figure}[h]
\includegraphics[width=0.99\columnwidth]{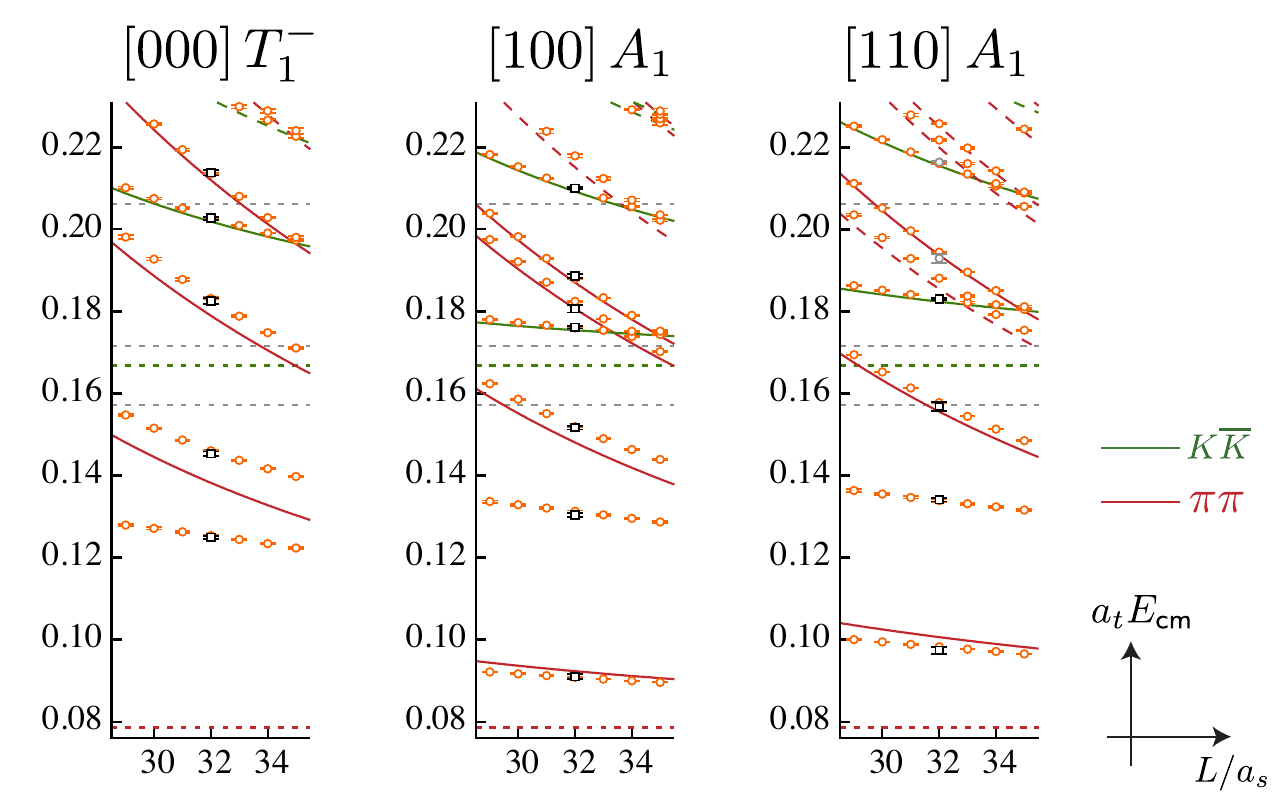}
\vspace{-0.3cm}
\caption{As Figure~\ref{spec_T1_A1} with the addition of orange points with errorbars showing the spectrum corresponding to the parameterization in Eq.~\ref{coupled_fit}.}\label{spec_recon}
\end{figure}

We make use of a total of 34 energy levels, shown by the black points in Figures~\ref{spec_T1_A1} and \ref{spec_E_B}. Four of these states show a significant overlap with a $\kk$ operator, whilst the remaining levels in the coupled-channel region dominantly overlap with $\pi\pi$ operators. This corresponds to using all energy levels below $a_tE_\cm=0.22$, or below the the first unknown ``$\pi\pi$'' level, whichever is lowest. This spectrum can be described by the $K$-matrix of Eq.~\ref{eq_km_cc}, with $N=0$, with parameters
\begin{widetext}
\begin{center}
\begin{tabular}{rll}
$m =$                         & $ 0.13170(36)(6) \cdot a_t^{-1}$   & 
\multirow{5}{*}{ $\begin{bmatrix*}[r] 1 & -0.20 & -0.24 & -0.27 &  0.08 &  0.10 \\ 
                                    	&  1    & -0.77 &  0.69 & -0.19 & -0.67 \\
                                    	&       & 1     & -0.58 &  0.40 &  0.90 \\
                                    	&       &       &  1    & -0.03 &  0.39 \\
                                    	&       &       &       &  1    &  0.53 \\
                                    	&       &       &       &       &  1    \end{bmatrix*}$ } \\
$g_{\pi\pi} =$                                & $0.4463(80)(40)$  & \\
$g_{\kksm} =$                        & $0.71(11)(134)$   & \\
$\gamma_{\pi \pi,\,\pi \pi} = $               & $1.56(94)(30)  \cdot a_t^{-2}$   & \\
$\gamma_{\pi \pi,\,\kksm} = $        & $6.7(26)(143) \cdot a_t^{-2}$& \\
$\gamma_{\kksm,\,\kksm} = $ & $6.8(56)(27) \cdot a_t^{-2}$   & \\[1.3ex]
&\multicolumn{2}{l}{ $\chi^2/ N_\mathrm{dof} = \frac{38.2}{34-6} = 1.37 $\,.}
\end{tabular}
\end{center}
\vspace{-1.5cm}
\begin{equation} \label{coupled_fit}\end{equation}
\end{widetext}
We observe a quite reasonable description as measured by the $\chi^2/N_\mathrm{dof}$, noting however that some parameters are rather strongly correlated, suggesting there is some unnecessary freedom in Eq.~\ref{eq_km_cc}. The phase-shifts, $\delta_{\pi\pi}$, $\delta_{\kksm}$, and inelasticity, $\eta$, defined in
\begin{align}
t_{\pi\pi, \pi\pi} &= \frac{\eta e^{2 i \delta_{\pi\pi}} -1 }{2i \rho_{\pi\pi}} \nonumber\\
t_{\kksm, \kksm} &= \frac{\eta e^{2 i \delta_{\kksm}} -1 }{2i \rho_{\kksm}} \nonumber\\
t_{\pi\pi, \kksm} &= \frac{\sqrt{1-\eta^2} e^{i (\delta_{\pi\pi}+ \delta_{\kksm} ) }}{2 \sqrt{ \rho_{\pi\pi} \rho_{\kksm} } },
\end{align}
are presented in Figure~\ref{coupled_default_K}. We clearly observe the same resonant behavior in $\delta_{\pi\pi}$ in the elastic region that we saw previously. We further note that there is very little coupling between $\pi\pi$ and $\kk$ above $\kk$ threshold, and that the $\kk \to \kk$ amplitude shows signs of being mildly repulsive. That this amplitude describes the finite volume spectra rather well can be seen in Figure~\ref{spec_recon}.

Of course we should be careful not to draw too many conclusions from this first description -- we cannot be certain that our choice of parameterization has not forced certain features onto the result. To investigate this, we consider a range of parameterizations. For example we may vary the order of the $K$-matrix polynomial, $N$, in Eq.~\ref{eq_km_cc}. We may also consider implementing a ``running pole coupling'', where the factors $g_i$ in Eq.~\ref{eq_km_cc} are replaced with energy dependent polynomials, $g_i\to g_i(s)=\sum_{m=0}^M g_i^{(m)}\,s^m$, where all $g$'s are real constants. Another variation drops the Chew-Mandelstam part of the phase-space, instead just using the simple phase-space, $I_i(s) = -i\rho_i(s)$, which satisfies unitarity in a minimal way. Such a form is not ideal if we wish to extrapolate far below thresholds, as a kinematic singularity appears at $s=0$, but we will not have cause to go so far below threshold in this case. We summarize these variations in Table~\ref{tab_coupled_fits} showing the resulting $\chi^2/N_\mathrm{dof}$. Fits of comparable quality were found representing the elements of the inverse $K$-matrix as polynomials (as was used in Ref~\cite{Wilson:2014cna}),
\begin{align}
K^{-1}_{ij}=\sum_{m=0}^M c_{ij}^{(m)}s^m,
\end{align}
however they produced results with very high degrees of parameter correlation, leading to an unreliable estimate of statistical error, and we will not discuss them further.

\begin{figure}
\includegraphics[width=0.95\columnwidth]{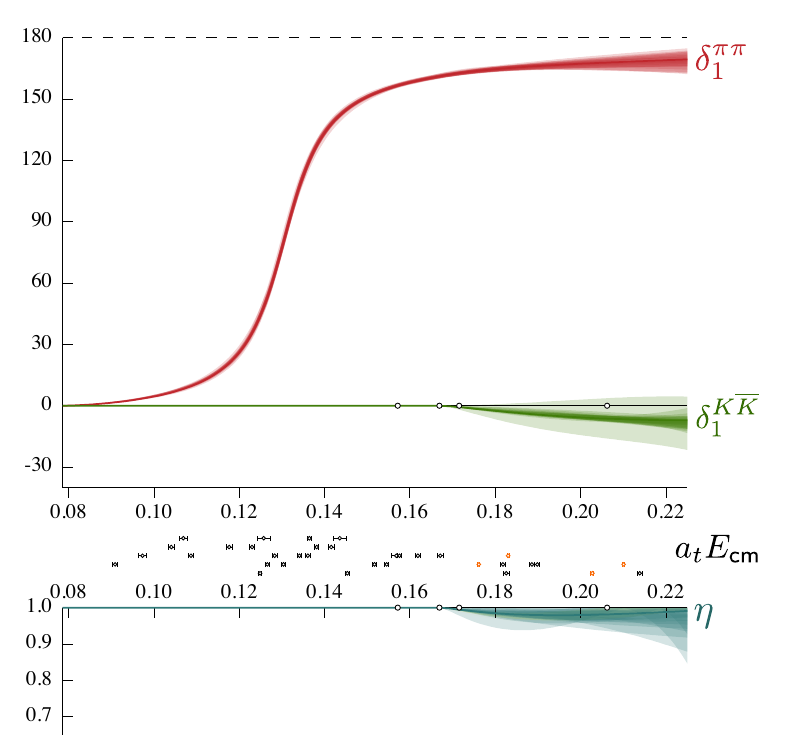}
\caption{As Figure~\ref{coupled_default_K} for a range of $K$-matrix parameterizations taken from the list given in Table~\ref{tab_coupled_fits}.}\label{coupled_variations}
\end{figure}

We plot the phase-shifts and inelasticities for a selection of fits presented in Table~\ref{tab_coupled_fits} in Fig.~\ref{coupled_variations} where we see that the lattice energy levels very tightly constrain each of these forms to give an amplitude description which does not vary significantly with parameterization.

\begin{table*}
\begin{ruledtabular}
\begin{tabular}{clcc}
Type          & Explicit form                                                                     & $N_\mathrm{pars}$ & $\chi^2/N_\mathrm{dof}$\\
\hline\hline\\[-1.2ex]
\multirow{9}{*}{$K\mathrm{-matrix}$}

              & $K_{ij} = \frac{g_i g_j}{m^2 -s} +\gamma^{(0)}_{ij}$;\quad $g_{\kksm}=0$                     & 5 & 1.75\\[0.6ex] 
              & $K_{ij} = \frac{g_i g_j}{m^2 -s} +\gamma^{(0)}_{ij}$;\quad $\gamma_{\pi\pi,\kksm}=0$         & 5 & 1.48\\ 
              & $K_{ij} = \frac{g_i g_j}{m^2 -s} +\gamma^{(0)}_{ij}$;\quad $\gamma_{\kksm,\kksm}=0$            & 5 & 1.37\\[0.6ex] 
              & $K_{ij} = \frac{g_i g_j}{m^2 -s} +\gamma^{(0)}_{ij}$                                  & 6 & 1.37\\[0.6ex] 
              & $K_{ij} = \frac{g_i g_j}{m^2 -s} +\gamma^{(1)}_{ij} s$                                & 6 & 1.41\\[0.6ex] 
              & $K_{ij} = \frac{g_i g_j}{m^2 -s} +\gamma^{(0)}_{ij}+\gamma_{ij}^{(1)} s$;\quad $g_{\kksm}=0$  & 8 & 1.52\\[0.6ex] 
              & $K_{ij} = \frac{g_i g_j}{m^2 -s} +\gamma^{(0)}_{ij}+\gamma_{ij}^{(1)} s$               & 9 & 1.39\\[1ex] 
\hline\hline\\[-1ex]
\multirow{12}{*}{
$\begin{matrix}
K\mathrm{-matrix} \\
\mathrm{with}\, g(s)
\end{matrix}$
}

              & $K_{ij} = \frac{g_i(s) g_j(s)}{m^2 -s} +\gamma^{(0)}_{ij}$;   & \multirow{2}{*}{6} & \multirow{2}{*}{1.34}\\[0.5ex] 
              & $\quad\quad g_{i}(s)= g_{i}^{(0)}+g_{i}^{(1)} \, s$;\quad $\gamma_{\kksm,\kksm}=0,\; \gamma_{\pi\pi,\kksm}=0$ & \\[1ex]
              & $K_{ij} = \frac{g_i(s) g_j(s)}{m^2 -s} +\gamma^{(0)}_{ij}$;   & \multirow{2}{*}{6} & \multirow{2}{*}{1.33}\\[0.5ex] 
              & $\quad\quad g_{i}(s)= g_{i}^{(0)}+g_{i}^{(1)} \, s$;\quad $\gamma_{\pi\pi,\pi\pi}=0,\; \gamma_{\pi\pi,\kksm}=0$ & \\[1ex]
              & $K_{ij} = \frac{g_i(s) g_j(s)}{m^2 -s} +\gamma^{(0)}_{ij}$;   & \multirow{2}{*}{7} & \multirow{2}{*}{1.38}\\[0.5ex] 
              & $\quad\quad g_{\pi\pi}(s)= g_{\pi\pi}^{(0)} + g_{\pi\pi}^{(1)} \, s$,\; $g_{\kksm}(s)=g_{\kksm}^{(0)}$ & \\[1ex]
              & $K_{ij} = \frac{g_i(s) g_j(s)}{m^2 -s} +\gamma^{(0)}_{ij}$;   & \multirow{2}{*}{7} & \multirow{2}{*}{1.35}\\[0.5ex] 
              & $\quad\quad g_{\pi\pi}(s)= g_{\pi\pi}^{(0)}$,\; $g_{\kk}(s)=g_{\kksm}^{(0)} + g_{\kksm}^{(1)}\,s$ & \\[1ex]
              & $K_{ij} = \frac{g_i(s) g_j(s)}{m^2 -s} +\gamma^{(0)}_{ij}$;   & \multirow{2}{*}{8} & \multirow{2}{*}{1.37}\\[0.5ex] 
              & $\quad\quad g_{i}(s)= g_{i}^{(0)}+g_{i}^{(1)}\, s$ & \\[1ex]
\hline\hline\\[-1ex]
\multirow{5}{*}{
$\begin{matrix}
K\mathrm{-matrix} \\
\mathrm{with}\\
 I_i(s)=-i\rho_i(s)
\end{matrix}$
}
              & $K_{ij} = \frac{g_i g_j}{m^2 -s} +\gamma^{(0)}_{ij}$;\quad $g_{\kk}=0$              & 5 & 1.57\\[0.6ex]
              & $K_{ij} = \frac{g_i g_j}{m^2 -s} +\gamma^{(0)}_{ij}$;\quad $\gamma_{\pi\pi,\kksm}=0$  & 5 & 1.40\\[0.6ex] 
              & $K_{ij} = \frac{g_i g_j}{m^2 -s} +\gamma^{(0)}_{ij}$;\quad $\gamma_{\kksm,\kksm}=0$     & 5 & 1.58\\[0.6ex]
              & $K_{ij} = \frac{g_i g_j}{m^2 -s} +\gamma^{(0)}_{ij}$                           & 6 & 1.45\\[0.6ex]
\end{tabular}
\end{ruledtabular}
\caption{Coupled-channel $K$-matrix parameterizations.}
\label{tab_coupled_fits}
\end{table*}

\newpage
\section{resonance interpretation} \label{res_interp}

Although we constrain partial-wave $t$-matrices only for real values of energy, either from experimental scattering, or in this case from finite-volume spectra, the amplitudes may be considered to be functions of a complex value of $s= E_\mathsf{cm}^2$. That the singularity structure of $t(s)$ might be important is already apparent if we consider the elastic unitarity condition, $\mathrm{Im}\,  t(s) = \rho(s)\, |t(s)|^2$, where $\rho(s) = 2 k_\mathsf{cm}(s)/\sqrt{s}$ has a square root branch cut beginning at the kinematic threshold. It follows that $t(s)$ also has this branch cut and remains single-valued only if we consider two Riemann sheets, one where $\mathrm{Im}\, k_\mathsf{cm}$ is positive, the ``physical'' sheet, named because physical scattering corresponds to energies $s+i\epsilon$ on this sheet, and one where $\mathrm{Im}\, k_\mathsf{cm}$ is negative, the ``unphysical'' sheet. As more two-body channels open, a greater multiplicity of sheets arises, corresponding to the increased number of channel momenta.

The rapid phase and amplitude variation that we associate with a narrow resonance can be caused by a pole at complex values of $s = s_0 =  \left(m - i \tfrac{1}{2}\Gamma \right)^2$ on unphysical sheets\footnote{a conjugate pole must also be present at $s_0^*$, but this pole is usually much further from the region of physical scattering.}. We may consider our parameterized $t$-matrices, looking for poles at complex values of $s$, of the form $t_{ij}(s)\sim\frac{c_i c_j}{s_0 - s}$ where we factorize the residue of the pole into couplings to each channel, $i$. 

\begin{figure}[b]
\includegraphics[width=0.95\columnwidth]{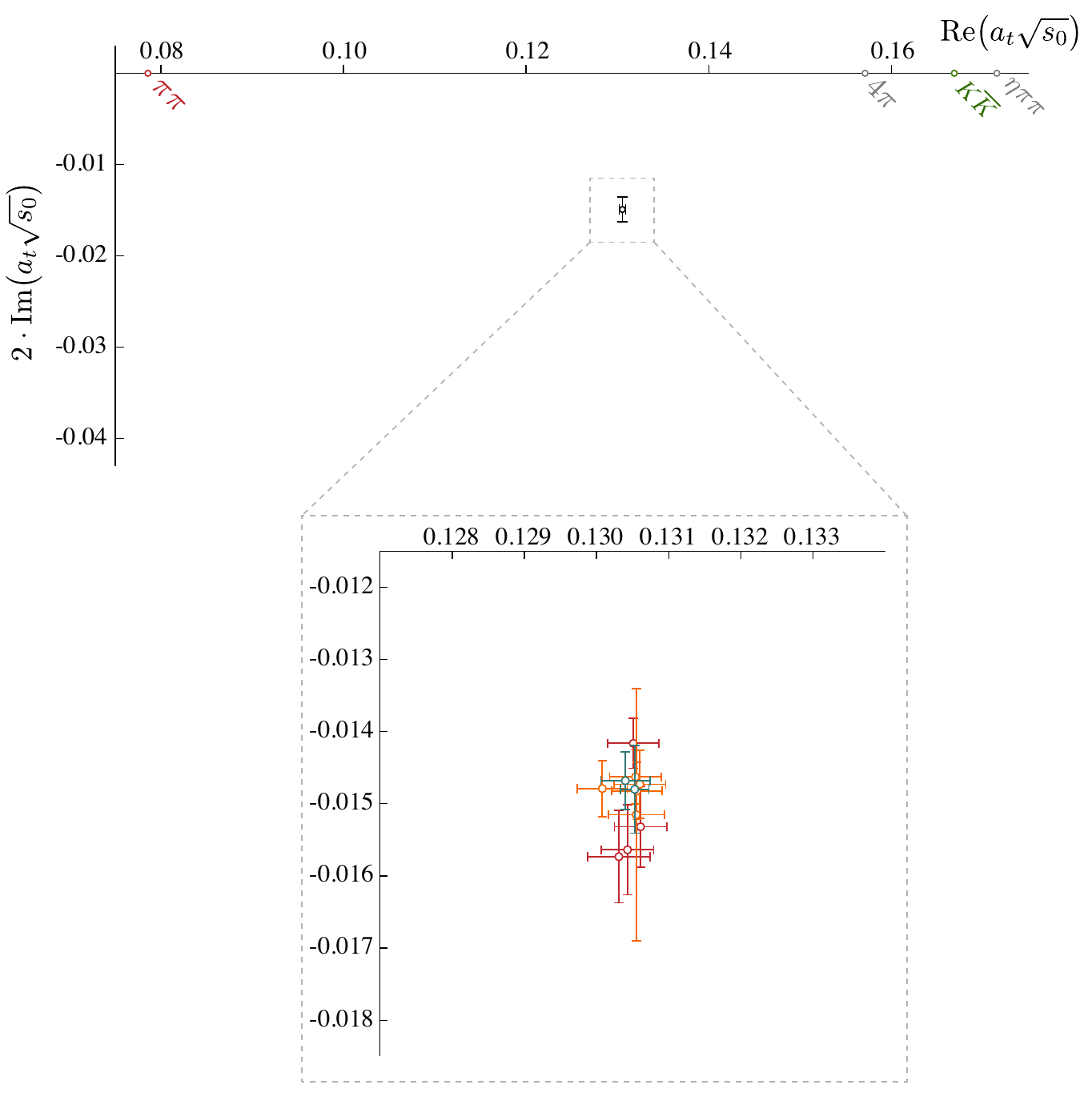}
\caption{Resonance pole position on first unphysical sheet. Zoomed region shows the pole found for a variety of parameterizations: various descriptions of the elastic amplitude (red), various $K$-matrix descriptions of the coupled-channel $t$-matrix, using the Chew-Mandelstam phase-space (orange), and using the simple phase-space (blue).  }\label{pole_position}
\end{figure}

We find that in every case we considered capable of describing the finite-volume spectrum, both in single-channel and coupled-channels, there is a statistically well-determined pole near $a_t\sqrt{s_0} =\left(0.1306-\frac{i}{2}0.015\right)$. Parameterizations that do not contain the freedom for a resonance pole to occur were not capable of successfully describing the finite volume spectra. Figure~\ref{pole_position} illustrates the position of the found pole, with the lower portion of the diagram showing a zoomed region in which the determined pole is shown for a range of different parameterizations. A best estimate for the pole position, whose uncertainties allow for the spread over parameterizations is 
\begin{align}
a_t\sqrt{s_0}&=\left(0.13055(36)-\frac{i}{2}0.0150(14)\right). \nonumber
\end{align}
The corresponding coupling to the $\pi\pi$ channel also shows very little variation under parameterizations with a good estimate being $a_t\, c_{\pi\pi} = 0.049(3)\, e^{-i \pi \, 0.06(1)}$. The coupling to $K\overline{K}$, which only arises in coupled-channel analysis is not well determined, having a large statistical uncertainty. Along with the observation that the elastic data can be very well described without invoking any $K\overline{K}$ amplitude, we conclude that we have not reliably constrained $c_{\kk}$. This is to be expected as the effect of the $K\overline{K}$ amplitude on the spectrum in finite-volume decays exponentially as we go lower in energy below the $K\overline{K}$ threshold.

If we follow the procedure used in previous calculations, making use of the $\Omega$ baryon mass determined on these lattice configurations, to set a physical scale we find $a_t=\frac{a_t m_\Omega}{m_\Omega^{\mathrm{phys}}}$, where $a_t m_\Omega$ is determined using lattice QCD computation and $m_\Omega^{\mathrm{phys}} = 1672.5\,  \mathrm{MeV}$ is the experimental mass. Using 16 distillation vectors on this lattice we have determined $a_t m_\Omega=0.2789(16)$, which leads to an approximate pion mass of $m_\pi=236$ MeV. 

\begin{figure}
\includegraphics[width=0.99\columnwidth]{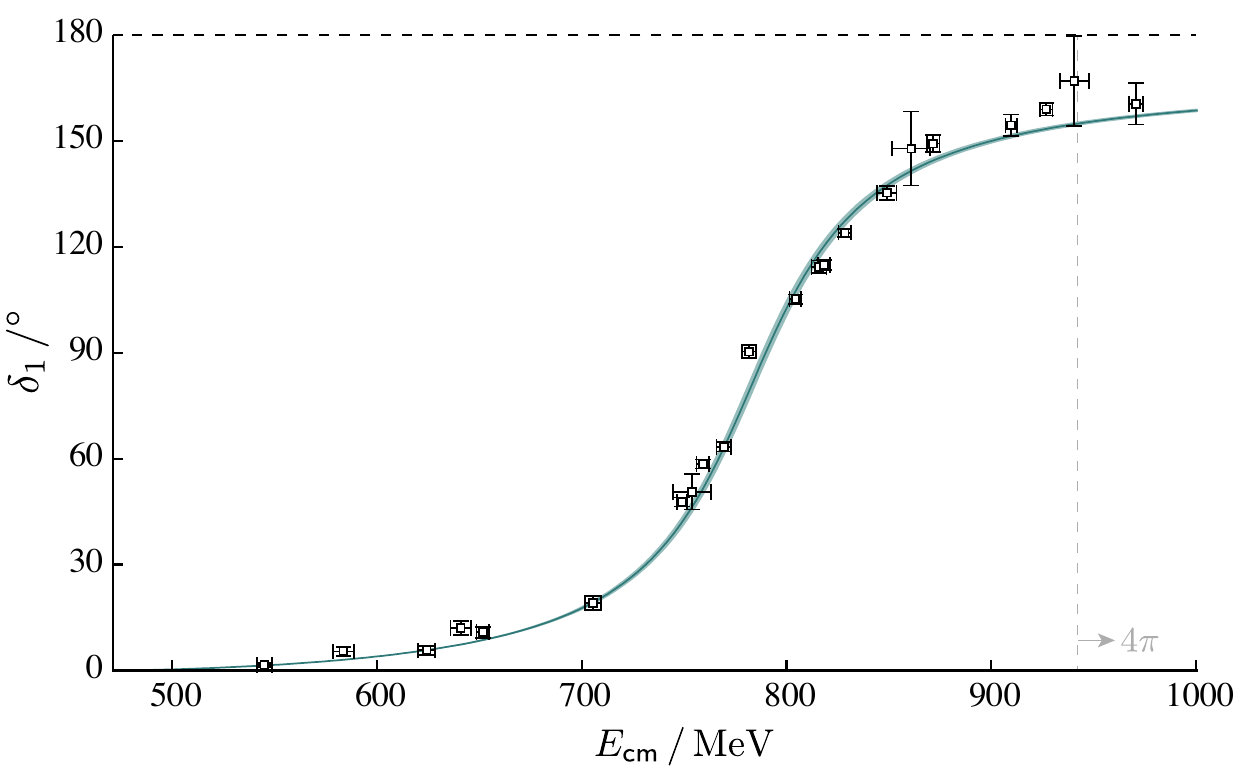}
\caption{Elastic Breit-Wigner fit of Figure~\ref{elastic_points_BW} with the lattice spacing set as described in the text.\label{BW_scale_set}}
\end{figure}

With this scale setting, the resonance pole is located at $\sqrt{s_0}=\left[   783(2)-\frac{i}{2}90(8)  \right] \mathrm{MeV}$. The scale-set Breit-Wigner mass and width of Eq.~\ref{bw_par_values} are $m_\mathrm{BW} = 790(2) \,\mathrm{MeV}$, $\Gamma_\mathrm{BW} = 87(2)\, \mathrm{MeV}$, and a plot of the corresponding phase-shift with the scale-set energy is presented in Figure~\ref{BW_scale_set}.

\begin{figure}
\includegraphics[width=0.99\columnwidth]{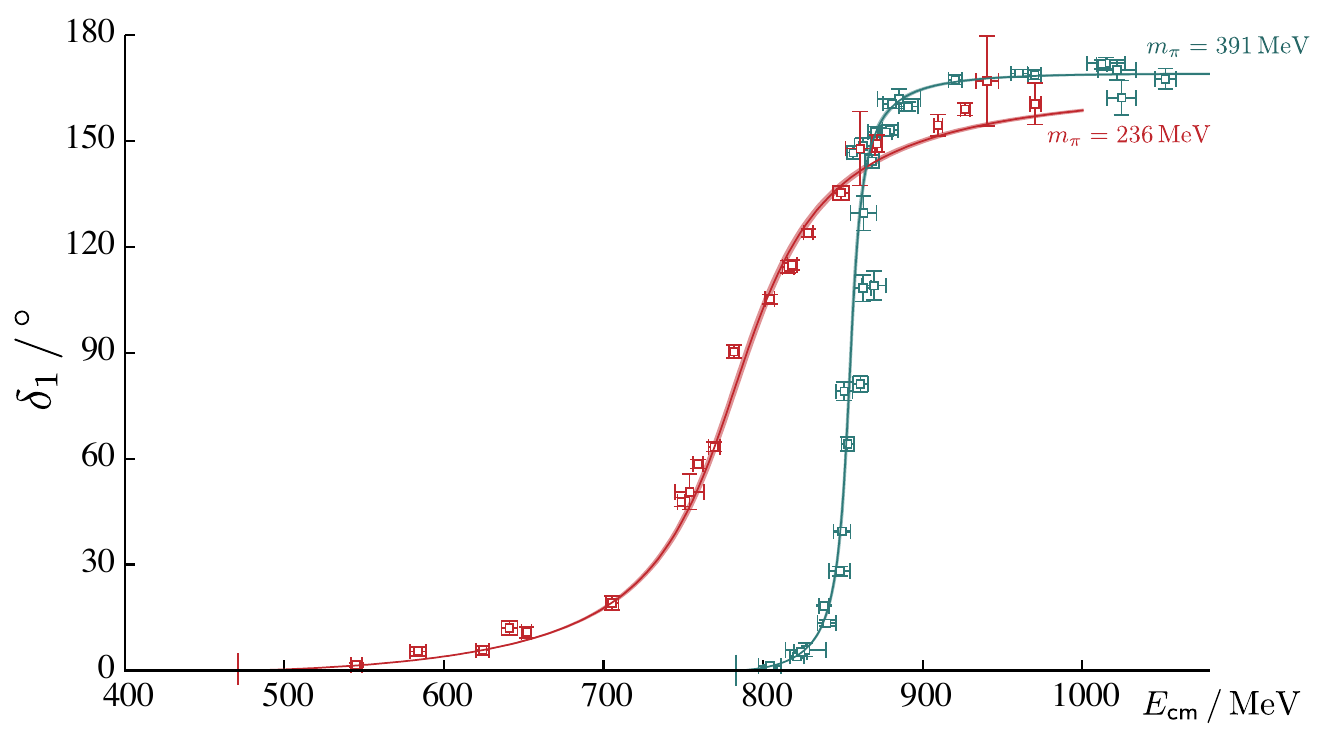}
\caption{Elastic $P$-wave $\pi\pi$ isospin=1 phase shifts for pions of mass $\sim 391$ MeV (blue) and $\sim 236$ MeV (red). Curves correspond to two parameter Breit-Wigner fits as described in the text and summarized in Table~\ref{bw_par_values_compared}.}\label{fig_840_860}
\end{figure}

An earlier calculation by the Hadron Spectrum Collaboration considered elastic $\pi\pi$ scattering using lattice configurations with the same quark and gluon action, but with a larger value of the $u,d$ quark mass, such that the pion had a mass of $391\,\mathrm{MeV}$~\cite{Dudek:2012xn}. We compare the Breit-Wigner parameters in this study with those determined for $m_\pi = 391 \,\mathrm{MeV}$ in Table~\ref{bw_par_values_compared}\footnote{The results presented in \cite{Dudek:2012xn} suffer from a small error in the computation of the off-diagonal data covariance, which we fix here, leading to a very small shift (at the level of $1\sigma$) in the quoted Breit-Wigner parameters with respect to that reference.}. The corresponding phase-shifts are shown in Fig.~\ref{fig_840_860}. We note that the Breit-Wigner couplings $g_R$ show good agreement between the two different quark masses, as has been suggested in unitarized versions of chiral perturbation theory~\cite{Nebreda:2010wv}. The apparent difference in the pole residue coupling, $c_{\pi\pi}$, is completely explicable in terms of the $P$-wave barrier -- since the {$t$-matrix} near threshold must behave like $t(s) \sim k^2$, we may consider $c_{\pi\pi} = \tilde{c}_{\pi\pi} \, k(s_0)$ where $k(s_0)$ is the $\mathsf{cm}$-frame momentum at the pole position. It follows that
\vspace{-0.4cm}
\begin{align*}
	\left| \frac{c_{\pi\pi}^\mathsf{391}}{c_{\pi\pi}^\mathsf{236}} \right| &= 0.56(2)  \\
	&=  \left|\frac{\tilde{c}_{\pi\pi}^\mathsf{391}}{\tilde{c}_{\pi\pi}^\mathsf{236}} \right| \left| \frac{k^\mathsf{391}}{k^\mathsf{236}} \right|  \\
	&=  \left| \frac{\tilde{c}_{\pi\pi}^\mathsf{391}}{\tilde{c}_{\pi\pi}^\mathsf{236}} \right|  \left| \frac{173 - 7.7 i}{312 - 26.6 i} \right| \\
	&=  \left| \frac{\tilde{c}_{\pi\pi}^\mathsf{391}}{\tilde{c}_{\pi\pi}^\mathsf{236}} \right|\,  0.552, 
\end{align*}
and thus $|\tilde{c}_{\pi\pi}^\mathsf{391}| \approx |\tilde{c}_{\pi\pi}^\mathsf{236}|$.

\begin{table}
\begin{ruledtabular}
\begin{tabular}{rrr}
                         & This work & Ref.~\cite{Dudek:2012xn}\\
\hline\hline\\[-1.0ex]
$a_t m_\pi$              &  $0.03928(18)$                & $0.06906(13)$ \\
$a_t m_R$               &  $0.13175(35)(5)$  			& $0.15095(14)(4)$  \\
$g_R$                   &  $5.688(70)(26)$  			& $ 5.698(97)(3)$  \\[1.4ex]
\hline\hline\\[-1.0ex]
$m_\pi$             &  $236(2)$ MeV  & $391(1)$ MeV\\
$m_R$               &  $790(2)$ MeV  & $855(1)$ MeV \\[1.4ex]
\hline\hline\\[-1.0ex]
$\mathrm{Re}(\sqrt{s_0})$ 	& $783(2)$ MeV & $853(2)$ MeV \\ 
$-2 \, \mathrm{Im}(\sqrt{s_0})$    & $85(2)$ MeV & $ 12.4(6)$ MeV \\ 
$|c_{\pi\pi}|$            & $ 288(4)$ MeV & $ 162(4)$ MeV \\ 
$\mathrm{Arg}(c_{\pi\pi})$
                          & $ -0.059(1)\,\pi$     & $-0.033(1) \,\pi$ \\[1.4ex]
\hline\hline\\[-1.0ex]
$\chi^2/ N_\mathrm{dof}$ & $\frac{24.9}{22 - 2} = 1.25$ & $\frac{28.7}{31 - 2}= 0.98$ \\[1.2ex]
\end{tabular}
\end{ruledtabular}
\label{bw_par_values_compared}
\caption{
A comparison of the results of this study and ref.~\cite{Dudek:2012xn}. These numbers compare the Breit-Wigner description only and the quoted pole is from that single parameterization. }
\end{table}

\section{Summary} \label{conclusion}

The $\rho$ resonance was extracted using a detailed spectrum of lattice QCD energy levels working in a $(\sim 4\, \mathrm{fm})^3$ cubic volume with a pion mass of $236$ MeV. Using the variational method and a large diverse basis of operators, energy levels were obtained in the elastic $\pi\pi$ scattering region and the near-threshold coupled-channel $\pi\pi-K\overline{K}$ energy region and these were used to constrain the $I=1$ $J^P=1^-$ and $J^P=3^-$ scattering amplitudes. By making use of the formalism relating the elastic and coupled-channel scattering amplitudes to the spectrum of eigenstates in a finite volume, we were able to extract phase-shifts and inelasticity for the coupled $\pi\pi-K\overline{K}$ system. The elastic region was found to feature a narrow resonance, which persists when the coupled $\kk$ channel is also considered. A range of $t$-matrix parameterizations lead to consistent resonance parameters in the sense of a pole in the complex energy plane. Using the $\Omega$ baryon to set the scale this pole is located at $\sqrt{s}_\rho=\left(783(2)-\frac{i}{2}90(8)\right)$ MeV. A simple Breit-Wigner description of the elastic amplitude works well and gives a coupling that is consistent with the value determined at a larger pion mass and that extracted from experimental data.

In a coupled-channel analysis we found the $\pi\pi-K\overline{K}$ system to be only weakly coupled for the range of energies we considered, and only small phase-shifts were observed in the $K\overline{K}$ channel. This is only the second example of the extraction of a coupled-channel scattering matrix from lattice QCD, following the earlier study of $\pi K,\, \eta K$ ~\cite{Dudek:2014qha,Wilson:2014cna}. This first exploratory study above $K\overline{K}$ threshold neglected three- and four-hadron contributions that have been observed to be suppressed in experimental studies. Progress in the development of a finite-volume formalism capable of dealing with these higher-multiplicity channels is ongoing~\cite{Hansen:2013dla,Hansen:2014eka,Hansen:2015zga}. 

The consistency and broad applicability of these methods to extract resonance properties from lattice QCD is now being demonstrated with successful applications at multiple pion masses in various quantum numbers. Future studies will aim to shed light on longstanding mysteries such as the $a_0(980)$, $f_0(980)$ resonances, and investigate excited hadron states with exotic quantum numbers.


\begin{acknowledgments}

We thank our colleagues within the Hadron Spectrum Collaboration, and in particular, thank B\'alint Jo\'o for his help. 
The software codes
{\tt Chroma}~\cite{Edwards:2004sx}, {\tt QUDA}~\cite{Clark:2009wm,Babich:2010mu}, {\tt QPhiX}~\cite{ISC13Phi}, and {\tt QOPQDP}~\cite{Osborn:2010mb,Babich:2010qb} were used to compute the propagators required for this project. The contractions were performed on clusters at Jefferson Laboratory under the USQCD Initiative and the LQCD ARRA project. This research was supported in part under an ALCC award, and used resources of the Oak Ridge Leadership Computing Facility at the Oak Ridge National Laboratory, which is supported by the Office of Science of the U.S. Department of Energy under Contract No. DE-AC05-00OR22725.
This research is also part of the Blue Waters sustained-petascale computing project, which is supported by the National Science Foundation (awards OCI-0725070 and ACI-1238993) and the state of Illinois. Blue Waters is a joint effort of the University of Illinois at Urbana-Champaign and its National Center for Supercomputing Applications. This work is also part of the PRAC ``Lattice QCD on Blue Waters''. This research used resources of the National Energy Research Scientific Computing Center (NERSC), a DOE Office of Science User Facility supported by the Office of Science of the U.S. Department of Energy under Contract No. DE-AC02-05CH11231.
The authors acknowledge the Texas Advanced Computing Center (TACC) at The University of Texas at Austin for providing HPC resources that have contributed to the research results reported within this paper. Gauge configurations were generated using resources awarded from the U.S. Department of Energy INCITE program at Oak Ridge National Lab, and also resources awarded at NERSC. RAB, RGE and JJD acknowledge support from U.S. Department of Energy contract DE-AC05-06OR23177, under which Jefferson Science Associates, LLC, manages and operates Jefferson Laboratory. JJD acknowledges support from the U.S. Department of Energy Early Career award contract DE-SC0006765. CET acknowledges partial support from the U.K. Science and Technology Facilities Council [grant number ST/L000385/1].

\end{acknowledgments}

\appendix

%
\section{Scattering analyses with an incomplete spectrum} 
\label{app:single}

\begin{figure}[b]
\includegraphics[width=0.99\columnwidth]{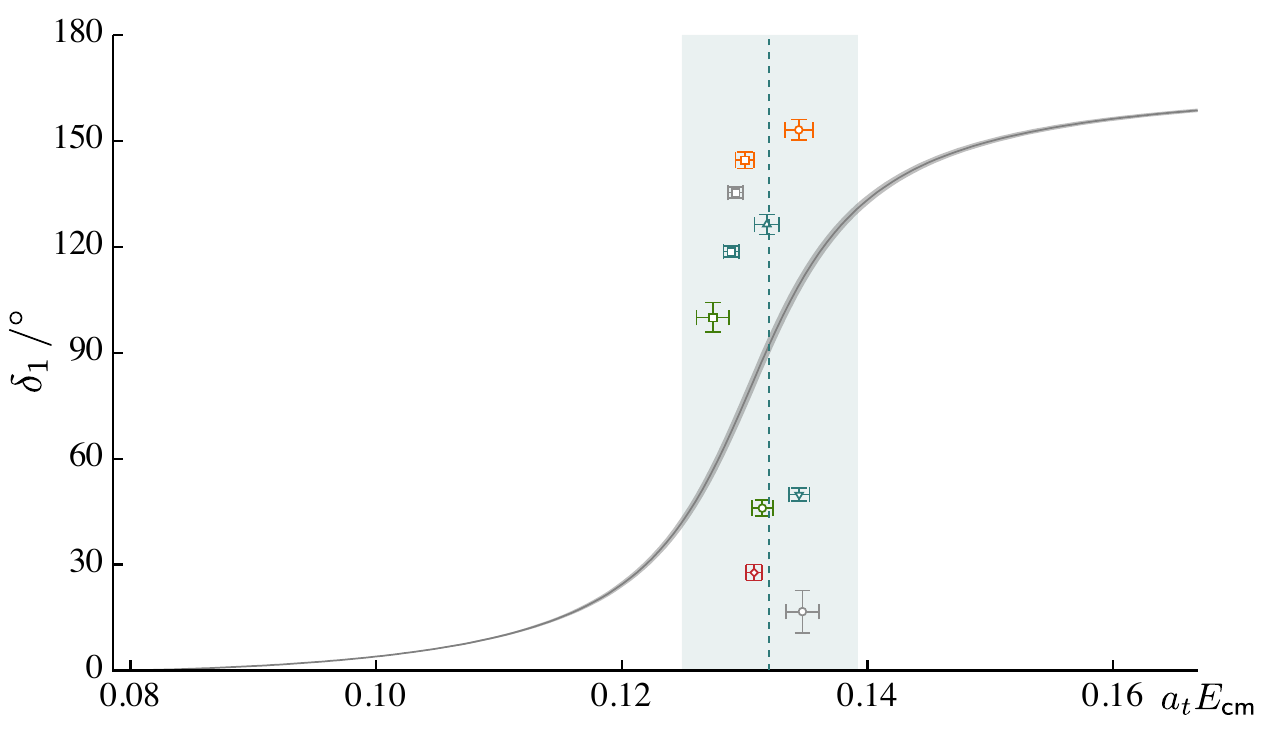}
\caption{Elastic phase-shift points (colors and symbols as in Figure~\ref{elastic_points}) corresponding to the spectrum extracted using only ``single-hadron-like'' operators, $\bar{\psi}\mathbf{\Gamma}\psi$. The grey band shows the elastic Breit-Wigner amplitude determined from the full spectrum as described in the body of the manuscript. The blue dashed vertical line and band indicate the Breit-Wigner mass and width.\label{fig_bad_spec}}
\end{figure}

In our basis we included operators that were specifically constructed to resemble both the $\pi\pi$ and $\kk$ states one would expect to exist in the absence of meson-meson interactions, as well as fermion bilinears, $\bar{\psi}\mathbf{\Gamma}\psi$, which resemble $q\bar{q}$-like constructions. As is visible in Figure~\ref{op_basis}, the low-lying states overlap with both sets of operators. This is what we might expect for a system containing a resonance that is dominantly $q\bar{q}$, but coupled to the decay channel $\pi\pi$.

In order to gauge the importance of including $\pi\pi$-like operators in the basis, we perform variational determination of the spectrum using only the ``single-meson-like'' operators in each irrep. The energies so determined are converted to elastic phase-shifts and are plotted in Figure~\ref{fig_bad_spec}. For comparison, we also show the Breit-Wigner elastic phase-shift curve determined in the text. As we have previously suggested~\cite{Dudek:2013yja}, when a narrow resonance is present, using only ``single-hadron-like'' operators tends to provide energies which lie roughly within one hadronic width of the mass of the state. It is clear that this limited operator basis may be useful to suggest the {\it presence} of a narrow state, but it cannot determine the resonant properties of such a state.

\pagebreak
\section{Flavor structure of $K \overline{K}$ operators}
\label{app:ops}

Our ``single-meson-like" and isospin-1 ``$\pi\pi$-like'' operators have already been described in detail elsewhere \cite{Dudek:2009qf,Dudek:2010wm,Thomas:2011rh,Dudek:2012gj}.  The isospin-1 ``$K \overline{K}$-like" operators are constructed to have positive $G$-parity (corresponding to neutral states with negative $C$-parity).  For example, the $I_z = +1$ component is proportional to,
$$
\left( \bar{s} \mathbf{\Gamma}_1 u \right) \; \left( \bar{d} \mathbf{\Gamma}_2 s \right)
+ \hat{G}\left[ \left(\bar{s} \mathbf{\Gamma}_1 u \right) \; \left( \bar{d} \mathbf{\Gamma}_2 s \right) \right] \, ,
$$
where $\mathbf{\Gamma}_i$ encodes the spin, derivative and momentum structure of the operator.  Here $\hat{G}$ is the $G$-parity transformation, $
\hat{G}\left[ \bar{q}_1 \mathbf{\Gamma} q_2 \right]
 = C \, \hat{G}[q_2] \mathbf{\Gamma} \hat{G}[\bar{q}_1] \, ,
$
where $C$ is the $C$-parity of the underlying spin and derivative structure, i.e. that for a flavorless $\bar{q} \mathbf{\Gamma} q$ operator.
The $K\overline{K}$ operators are projected onto definite irreps of the relevant symmetry group by summing over relative momenta in exactly the same way as for $\pi\pi$ operators.


\bibliographystyle{apsrev4-1}
\bibliography{paper}

\end{document}